\documentclass[twocolumn,superscriptaddress,amsmath,amssymb,prl]{revtex4-1}

\newcommand{\Ibadai}{Graduate School of Science and Engineering, Ibaraki University, Mito, Ibaraki 310-8512, Japan}
\newcommand{\Kinken}{Institute for Materials Research, Tohoku University (IMR), Aoba-ku, Sendai 980-8577, Japan}
\newcommand{\IMSS}{Muon Science Laboratory, Institute of Materials Structure Science, High Energy Accelerator Research Organization (KEK), Tsukuba, Ibaraki 305-0801, Japan}
\newcommand{\Sokendai}{Graduate University for Advanced Studies, SOKENDAI}
\newcommand{\ISSP}{Institute for Solid State Physics, University of Tokyo, Kashiwa, Chiba 277-8581, Japan}
\newcommand{\Nagoya}{Department of Applied Physics, Graduate School of Engineering, Nagoya University, Chikusa-ku, Nagoya 464-8603, Japan}

\usepackage{txfonts}
\usepackage[dvipdfmx]{graphicx}
\usepackage{dcolumn} 
\usepackage{bm} 
\usepackage{multirow}
\usepackage{textcomp}
\usepackage{multibib}
\usepackage[hypertex,colorlinks=true,linkcolor=black,citecolor=blue,filecolor=blue,urlcolor=blue,setpagesize=false,nesting=true]{hyperref}

\begin{document}
\title{Nonmagnetic Ground State in RuO$_2$ Revealed by Muon Spin Rotation}

\author{M.~Hiraishi}\email{masatoshi.hiraishi.pn93@vc.ibaraki.ac.jp}\affiliation{\Ibadai}\affiliation{\IMSS}
\author{H.~Okabe}\affiliation{\Kinken}\affiliation{\IMSS}
\author{A.~Koda}\affiliation{\IMSS}\affiliation{\Sokendai}
\author{R.~Kadono}\affiliation{\IMSS}
\author{T. Muroi}\affiliation{\ISSP}
\author{D. Hirai}\affiliation{\Nagoya}
\author{Z. Hiroi}\affiliation{\ISSP}

\date{\today}

\begin{abstract}
 The magnetic ground state of single crystalline RuO$_2$ was investigated by the muon spin rotation/relaxation ($\mu$SR) experiment. The spin precession signal due to the spontaneous internal magnetic field $B_{\rm loc}$, which is expected in the magnetically ordered phase, was not observed in the temperature range 5--400~K. Muon sites were evaluated by first-principles calculations using dilute hydrogen simulating muon as pseudo-hydrogen, and $B_{\rm loc}$ was simulated for the antiferromagnetic (AFM) structures with a Ru magnetic moment $|{\bm m}_{\rm Ru}|\approx0.05\mu_\mathrm{B}$ suggested from diffraction experiments. As a result, the possibility was ruled out that muons are localized at sites where $B_{\rm loc}$ accidentally cancels. Conversely, assuming that the slow relaxation observed in $\mu$SR spectra was part of the precession signal, the upper limit for the magnitude of $|{\bm m}_{\rm Ru}|$ was estimated to be $4.8(2)\times10^{-4}\mu_{\rm B}$, which is significantly less than $0.05\mu_\mathrm{B}$. These results indicate that the AFM order, as reported, is unlikely to exist in the bulk crystal.
 \end{abstract}

\maketitle
Ruthenium oxide RuO$_2$ with rutile structure is a material that has been applied in a variety of fields due to its high catalytic activity and remarkable chemical stability~\cite{review2012}.
From the viewpoint of electronic properties, RuO$_2$ has been considered to be an ordinary Pauli paramagnetic (i.e., nonmagnetic) metal~\cite{Ryden_1970}. However, it has recently been reported to have a topological electronic structure~\cite{Zhang2017,Vedran2018}. Furthermore, based on neutron diffraction~\cite{ND_2017} and resonant X-ray scattering experiments~\cite{Rxs_2019}, RuO$_2$ has been claimed to exhibit antiferromagnetic (AFM) ordering with a high N\'eel temperature ($>300$~K) and a Ru magnetic moment size of $\sim$$0.05\mu_\mathrm{B}$.
The presence of AFM phase has generated interest in its potential application for spintronic devices~\cite{spintronics1,spintronics2}.
In addition, strain-induced superconductivity has also been reported~\cite{SC_2020,SC_2021}, thereby fueling interest in its detailed electronic properties.

Following the inference of AFM ordering in RuO$_2$, theoretical predictions and experimental results for various anomalies related to transport phenomena have been reported. Anomalous Hall effects associated with collinear AFM phase and the non-concentric symmetric position of the nonmagnetic oxygen have been predicted from theoretical studies~\cite{Science_2020_AHE_theory}, and experimental result supporting this prediction has been reported~\cite{Feng2022_AHE}.
In addition, spin current due to the spin-splitter effect generated in the AFM phase has been theoretically proposed~\cite{SST_theory_2021}, which has been followed by reports of experimental results in favor of the prediction~\cite{SST_exp2_2022,SST_exp3_2022,SST_exp1_2022}.
Occurrence of the chirality magneto-optical effect is also theoretically predicted~\cite{CCMO_21}. 

However, the reported size of Ru magnetic moments is close to the limit of sensitivity in neutron and X-ray diffraction experiments. Moreover, a recent theoretical study suggests that AFM ordering may be induced by hole doping due to Ru vacancies in RuO$_2$ which is intrinsically nonmagnetic~\cite{arxiv_mazin_23}. Thus, verification of the AFM phase with local magnetic probes that are complementary to diffraction experiment is highly required. Motivated by this situation, we have investigated the magnetic ground state of RuO$_2$ by the muon spin rotation/relaxation ($\mu$SR). The experimental results show that the spontaneous muon spin precession signal expected in the magnetically ordered phase is not observed in the temperature range of 5--400 K. The first-principles density functional theory (DFT) calculations of muon sites exclude the possibility that muons are localized at sites where the internal magnetic field exerted by the Ru atoms is canceled by chance. These results support the scenario that the bulk crystal RuO$_2$ is a nonmagnetic metal.

$\mu$SR is a magnetically sensitive probe in which spin-polarized muons ($\mu^+$) stopped in the sample directly infer the internal magnetic field $B_{\rm loc}$ at the interstitial sites via the spontaneous Larmor precession.
Muons are provided as a 100\%-spin-polarized beam by proton accelerator facilities, which enables $\mu$SR measurements in zero magnetic field. The sensitivity of $\mu$SR as a local probe has been demonstrated in the discovery of AFM ordering in the parent compounds of high-$T_\mathrm{c}$ cuprate superconductors~\cite{Nishida_1987,Uemura_1987,Luke:89}, as well as in the study of superconductivity~\cite{Appeli:87,Uemura:89,Sonier:00}.
In addition, $\mu$SR has various advantages: the muon implantation energy is high enough ($\ge4$~MeV) to be surface independent (bulk sensitive), and the implanted muons (volume concentration $\sim$$10^5$ cm$^{-3}$) decay with an average lifetime of 2.2~\textmu s, so they do not accumulate in the sample unlike other ion beam irradiation measurements.
Therefore, the detection of $B_{\rm loc}$ due to ordered Ru moments by $\mu$SR could provide direct experimental evidence for the occurrence of AFM ordering.

For the quantitative assessment of $B_{\rm loc}$ using $\mu$SR, it is necessary to obtain information on the local electronic structure of $\mu^+$ in RuO$_2$; it behaves as a light isotope of proton/hydrogen (pseudo-H, hereafter denoted by the element symbol ``Mu") in matter.
To this end, first-principles DFT calculations were performed using Quantum Espresso~\cite{QE_2009,QE_2017,QE_2020_GPU} to investigate the Mu stopping sites (DFT+Mu). Structural relaxation calculations with H (to mimic Mu) using the GGA-PBE exchange correlation function were performed on a $3\times3\times3$ superlattice (54 Ru, 108 oxygen and H) of a rutile structure ($P4_2/mnm$) as the initial structure with cutoff energy of 60~Ry and $K$ points were set to $3\times3\times4$. Structure was relaxed until the maximum force on each atom was less than $1\times10^{-3}$~(Ry/Bohr) in the nonmagnetic state. For the on-site Coulomb interaction $U$ between Ru $4d$ orbitals, calculations were performed for values of $U = 0$--3 eV and results were found to be mostly independent of $U$. We then used the result for $U=2.0$ eV (where the AFM state is theoretically favored~\cite{ND_2017,Science_2020_AHE_theory,SC_2021}) for the simulation of $B_{\rm loc}$. More detailed information can be found in the Supplemental Material (SM)~\cite{sm}, which includes Refs.~\cite{fullprof,olex2_1,olex2_2,Shelxl,PP,ONCV}.

Single crystals of RuO$_2$ were grown from a polycrystalline sample in oxygen flow. RuO$_2$ reagent (Rare Metallic Co., Ltd., 99.95\%) was placed in the high temperature section of the furnace (1100$^\circ$C) and held for two weeks while oxygen flowed at a rate of 30 ml/min to obtain blue-black metallic single crystals in the low-temperature section downstream of the furnace. The crystals were about $3\times3\times0.5$ mm$^3$ at the maximum with a thin rhombic columnar shape. To evaluate the quality of these crystals, single-crystal (sc) XRD analysis and electrical resistivity measurements were performed on the samples from the same batch (for more details, see SM~\cite{sm}). The Ru occupancy was deduced to be 1.02(2) from sc XRD, indicating that the crystal is free of Ru defects at the percent level. The resistivity measurements showed a maximum residual resistivity ratio (rrr) of $\sim$1500. This is much higher than previously reported values ($\sim$30 \cite{SC_2021} or $\sim$200 \cite{Rogers:69}), confirming that the crystals were of extremely high quality. These crystals were loaded on a silver holder covering about $15\times15$ mm$^2$ in a mosaic pattern (corresponding to the powder average for the $\mu$SR spectra). As a reference, $\mu$SR measurements on RuO$_2$ powder (same supplier, 99.9\%) as delivered were performed, and obtained qualitatively similar results to those of the single-crystalline samples. However, the results showed certain differences in details presumably attributed to the sample qualities such as impurities and defects, whose details are found in SM~\cite{sm}.

Conventional $\mu$SR measurements were performed in zero field (ZF), weak longitudinal field (LF) and weak transverse field (TF) using the S1 instrument (ARTEMIS)~\cite{ARTEMIS}, a general-purpose $\mu$SR spectrometer installed at the S1 area of the Materials and Life science Experiment Facility (MLF), J-PARC. A 100\% spin-polarized $\mu^+$ beam ($\sim$4~MeV) was stopped in the sample mounted on the He gas-flow cryostat for controlling sample temperature. The time-dependent $\mu^+$ spin polarization, which reflects the magnetic field distribution at the Mu site, was monitored via the forward-backward asymmetry [$A(t)$] of positrons.
The $\mu$SR spectra were analyzed by least-square curve fits using ``musrfit"~\cite{musrfit}.
The background contribution which missed the sample was calibrated using a Ho sample of the same cross section as RuO$_2$ to be $A_\mathrm{BG}=0.0815$, and subtracted from $A(t)$.
The evaluation of $B_{\rm loc}$ due to Ru magnetic moments and putative Ru $4d$ ordered moments were performed using ``dipelec" code~\cite{dipelec}.

\begin{figure}[t]
  \centering
	\includegraphics[width=0.75\linewidth,clip]{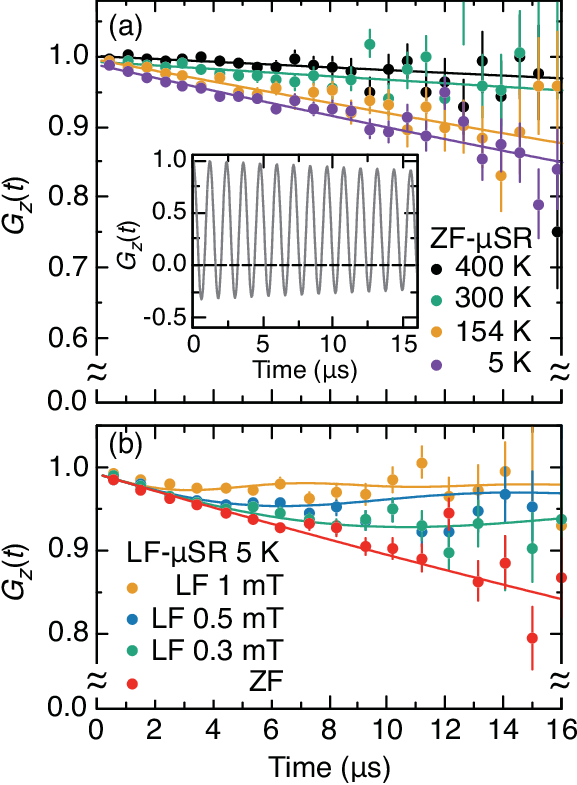}
	\caption{(a) Examples of ZF-$\mu$SR spectra at typical temperatures, and (b) magnetic field dependence of LF-$\mu$SR spectra at 5~K observed in single-crystalline RuO$_2$. Solid curves represent the least-square fit by Eq.~(\ref{Eq:ana}). Inset: the spontaneous spin precession signal expected in the presence of the presumed AFM order (see text).}
	\label{fig:spectra}
\end{figure}

Figure~\ref{fig:spectra}(a) shows ZF-$\mu$SR time spectra which are normalized by the asymmetry at $t=0$ to yield time-dependent spin polarization, $G_z(t)=A(t)/A_0$ [with $A_0=A(0)$]. The line shape of these spectra is well represented by slow exponential damping over the entire temperature range of 5--400~K, and no sinusoidal component for the Larmor precession is observed. This indicates that there is no homogeneous $B_{\rm loc}$ established at the Mu site.
Moreover, as shown in Fig.~\ref{fig:spectra}(b), the slow relaxation observed in ZF is completely suppressed by applying LF of 1~mT, indicating that $B_\mathrm{loc}$ is quasistatic.

Consequently, the curve fits of the spectra was performed using the following function,
\begin{align}
  A_0G_z(t)&=A_1G_\mathrm{KT}^\mathrm{L}(\lambda,B_\mathrm{LF},t),\label{Eq:ana}\\
  G_\mathrm{KT}^\mathrm{L}(\lambda,0,t)&=\frac{1}{3}+\frac{2}{3}(1-\lambda t)\exp(-\lambda t),\nonumber
\end{align}
where $G_\mathrm{KT}^\mathrm{L}(\lambda,B_\mathrm{LF},t)$ is the relaxation function for a static-Lorentzian field distribution with $\lambda$ being the relaxation rate~\cite{Uemura_1985}, $B_\mathrm{LF}$ is the external magnetic field.

The result of curve fits using Eq.~(\ref{Eq:ana}) is shown for the ZF time spectra at various temperatures and for the LF spectra at 5~K (including their $B_{\rm LF}$ dependence) in Fig.~\ref{fig:spectra}(a) and (b), both of which show good agreement. Considering that spin relaxation due to nuclear magnetic moments is usually well described by Gaussian Kubo-Toyabe functions~\cite{KT}, the origin of the observed Lorentzian-like behavior is unknown at this stage. Although the issue remains to be addressed in future work, these results indicate that $\lambda$ is not subject to fluctuation by self-diffusion of Mu nor magnetic fluctuation, ruling out the possibility that $B_{\rm loc}$ is effectively averaged out to zero due to the motional narrowing effect (see also the discussion below).
\par
\begin{figure}[t]
  \centering
	\includegraphics[width=\linewidth,clip]{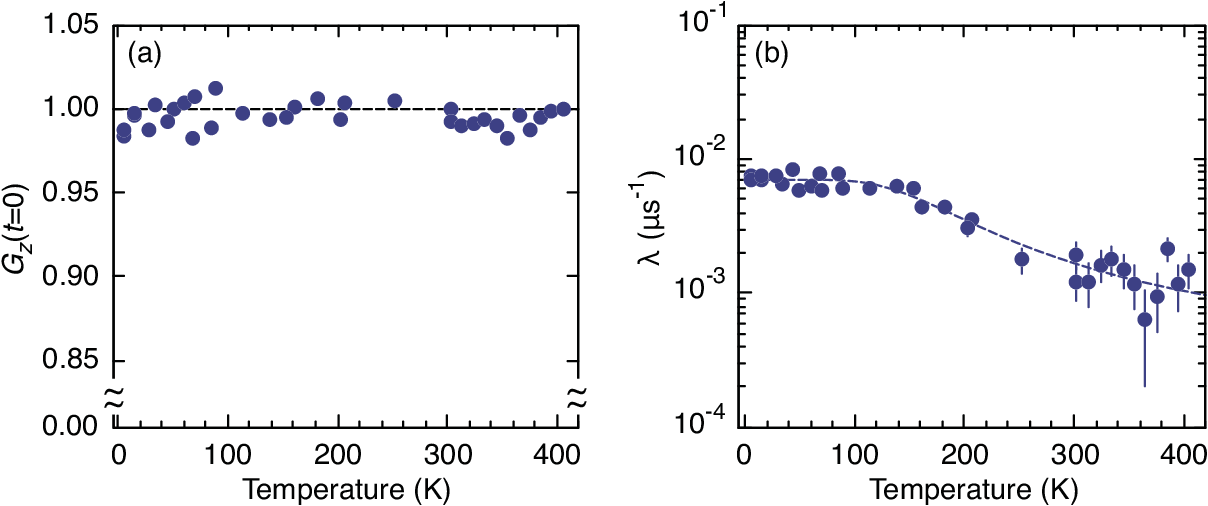}
	\caption{Temperature dependence of (a) the initial polarization and (b) relaxation rate obtained by the analysis using Eq.~(\ref{Eq:ana}). The dashed line in (b) is the result of a curve fit using Eq.~(\ref{Eq:lmd_ana}).}
	\label{fig:params}
\end{figure}

The temperature dependence of the initial Mu polarization [$G_z(0)$] is shown in Fig.~\ref{fig:params}(a), which is nearly unity irrespective of temperature for 5--400~K, indicating that there is no muonium (a neutral H atom-like state, Mu$^0$) formation nor fast depolarization components beyond the time resolution of the instrument.
Thus, implanted Mu are mostly in the diamagnetic state (Mu$^+$ or Mu$^-$).
Considering that RuO$_2$ is a metal, Mu is assumed to be an isolated state of Mu$^+$ due to shielding by conduction electrons.

As shown in Fig.~\ref{fig:params}(b), $\lambda$ is nearly constant at temperatures below about 100~K, while it decreases slowly at higher temperatures. Assuming that this is due to thermally activated diffusive motion of Mu, the temperature dependence of $\lambda$ is given as
\begin{align}
  \lambda(T)=\lambda_0\left[1+N\exp(-E_\mathrm{a}/k_\mathrm{B}T)\right]^{-1},
  \label{Eq:lmd_ana}
\end{align}
where $\lambda_0$ is the relaxation rate at $T\rightarrow0$~K, $E_\mathrm{a}$ is the activation energy, and $N$ is a physical quantities related to the density of states~\cite{Cox_06}.
 As shown by dashed line in Fig.~\ref{fig:params}(b), curve fit provides good agreement with the data, yielding $E_\mathrm{a}=62(8)$~meV being also typical for Mu diffusion in oxides~\cite{TUIto_23}.
Thus, it suggests that Mu diffusive motion may occur above $\sim$100 K, suggesting the possibility that $B_{\rm loc}$ is effectively reduced by the motional effect. Conversely, however, it suggests that at least below $\sim$100 K, $B_{\rm loc}$ is free from such motional effect.

In general, there are interstitial sites with high symmetry in antiferromagnets where $B_{\rm loc}$ exerted from magnetic ions cancels out.
Given the reported magnetic structure~\cite{ND_2017,Rxs_2019}, that is the case for the $8j$ site $(0.25, 0.25, 0.25)$ situated at the midpoint between Ru $(0, 0, 0)$ and Ru $(0.5, 0.5, 0.5)$, and the $4d$ site $(0, 0.5, 0.25)$. Provided that Mu accidentally occupies these sites, the spontaneous spin precession signal due to AFM ordering will be absent, thus insensitive to $B_{\rm loc}$. Therefore, it is important to gain reliable estimation of the Mu site(s) and associated sensitivity to $B_{\rm loc}$ expected for the reported magnetic structure.

\begin{figure}[tb]
  \centering
	\includegraphics[width=0.6\linewidth,clip]{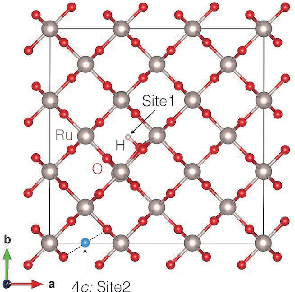}
	\caption{Local atomic configurations of hydrogen defect in RuO$_2$ obtained by DFT calculations with lattice relaxation in $3\times3\times3$ superlattice. Site1: the lowest energy site (forming OH bond). Site2 (light blue): the $c$-channel center ($4d$) site. The structure was displayed using VESTA~\cite{Vesta}.}
	\label{fig:h_strrlx}
\end{figure}

First, the most stable site (labeled Site1) for H inferred from structural relaxation calculation is shown in Fig.~\ref{fig:h_strrlx}. The H atom is about 0.1~nm from the nearest oxygen, and the bond direction is approximately in the $ab$ plane (there are four equivalent sites in the unit cell). This reproduces the previously reported result~\cite{RuO2_NEB_2021}, and is also in line with the general trend in oxides including rutile TiO$_2$~\cite{TiO2_NEB_2018} that Mu as pseudo-H forms OH bond with oxygen.
According to the total energy calculations performed with H fixed at the $8j$ and $4d$ sites, the total energies are about 2.1 and 1.4~eV higher than Site1, respectively, which rules out the possibility that Mu is stationary at these sites where $B_{\rm loc}$ cancels out.

The mass of Mu is approximately 1/9 that of H, and quantum effects such as zero-point energy $E_0$ can be pronounced. In particular, when $E_0$ is greater than the potential barrier $V_\mathrm{b}$ of the jump diffusion path to other stable sites, Mu may be delocalized to occupy site(s) away from Site1: such an effect has been actually reported for the iron-based superconductor LaFeAsO$_{0.49}$H$_{0.51}$~\cite{MH_PRB20}. According to earlier reports, the activation energy ($\approx V_\mathrm{b}$) of the diffusion path from the Site1 to the next OMu bonding site at the same $z$ in the same $c$-channel [$(x,y,z)\rightarrow(1-x,-y,z)$ in the unit cell] is estimated to be 0.27 or 0.217~eV (the latter includes the zero-point energy correction)~\cite{NEB_2015,RuO2_NEB_2021}. When $E_0$ of Mu (roughly three times that of H in the harmonic approximation) is greater than this barrier, Mu may be localized at the center in the $c$-channel. This site is the $4c$ site $(0.5, 0, 0)$ in the unit cell (denoted Site2, as shown in Fig.~\ref{fig:h_strrlx}).

Considering that the direction of the Ru moment ${\bm m}_\mathrm{Ru}$ is not uniquely provided in the literature~\cite{ND_2017,Rxs_2019} and that ${\bm m}_\mathrm{Ru}$ is theoretically predicted to vary from (001) to (100) as the electron filling progresses~\cite{Science_2020_AHE_theory}, $B_{\rm loc}$ is calculated for the arbitrary direction of ${\bm m}_\mathrm{Ru}$. Our calculations reveal that the $B_\mathrm{loc}$ is never canceled in any direction of ${\bm m}_\mathrm{Ru}$~(see SM~\cite{sm}).
In the case of ${\bm m}_\mathrm{Ru}\parallel (001)$, the corresponding $B_{\rm loc}$ is 6.17~mT for Site1 and 5.96~mT for Site2, yielding the precession frequencies 0.84 and 0.81~MHz for Site1 and Site2, respectively.
The precession signals for these frequencies should be readily observable with the present time resolution of J-PARC limited by the Nyquist frequency $1/2\tau\simeq6.25$~MHz which is determined by the FWHM ($\tau\approx80$~ns) of the pulsed $\mu^+$ beam.
For comparison, an expected $\mu$SR spectrum at Site1 (0.84~MHz) is shown in the inset of Fig.~\ref{fig:spectra}(a), where no such precession signal has been observed.

\begin{figure}[bt]
  \centering
  \includegraphics[width=0.7\columnwidth,clip]{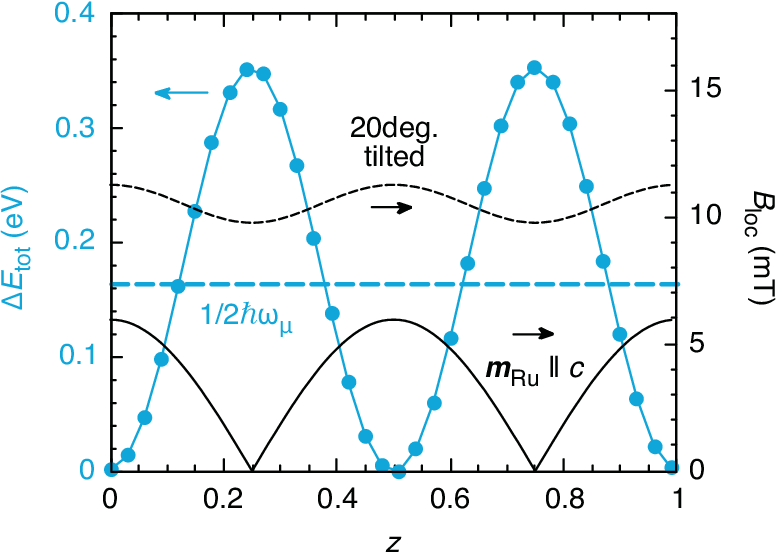}
  \caption{The total energy $\Delta E_\mathrm{tot}(\bm{r})$ and internal magnetic field $B_{\rm loc}(\bm{r})$ along the $\bm{r} =(0, 0.5, z)$ axis (corresponding to the $c$-channel center) calculated in the $2\times2\times3$ superlattice of RuO$_2$. The solid and dashed lines showing $B_{\rm loc}(\bm{r})$ (mapped on the right axis) are for the cases where the Ru moment ${\bm m}_\mathrm{Ru}$ is either parallel to (001) (the $c$-axis) or it is tilted 20$^\circ$ toward the (110) plane, respectively~\cite{Rxs_2019}. The horizontal dashed line is the Mu zero-point energy.}
  \label{Fig_Etot_jiba}
\end{figure}

Now, let us examine a more extreme possibility that Mu is delocalized along the $c$-channel center due to the zero-point energy. For the estimation of $E_0$ and $V_\mathrm{b}$, we investigated the total energy profile for $z$ position of H placed at the center of the $c$-channel in a $2\times2\times3$ superlattice, and evaluated its change $\Delta E_\mathrm{tot}$ with the following equation,
\begin{align}
  \Delta E_\mathrm{tot}(\bm{r})\equiv E_\mathrm{tot}(\bm{r})-E_\mathrm{min},
  \label{Eq_Etot}
\end{align}
where $\bm{r}=(0, 0.5, z)$ is the position vector in the unit cell and $E_\mathrm{min}$ is the minimum value in $E_\mathrm{tot}(\bm{r})$. The $z$ dependence of $\Delta E_\mathrm{tot}(\bm{r})$ and the magnitude of $B_{\rm loc}$ is shown in Fig.~\ref{Fig_Etot_jiba}. $\Delta E_\mathrm{tot}(\bm{r})$ exhibits oscillation with a $z/2$ period (shown on the left axis), with a maximum value of $V_\mathrm{b}=0.353$~eV. In the harmonic approximation of the periodic potential, the energy level splitting $\hbar\omega_\mu$ of Mu can be determined from the relation $\omega_\mu^2=2\pi^2V_\mathrm{b}/m_\mu d^2$~\cite{Yamaura_2019}, where $m_\mu=105.658$~MeV/$c^2$ is the $\mu^+$ mass, $d=c/2\simeq0.155$~nm is the length corresponding to the potential period. Then, $E_0=\tfrac{1}{2}\hbar\omega_\mu$ is estimated to be 0.163~eV, which is about half of $V_\mathrm{b}$. Therefore, even when Mu is at the center of the $c$-channel, it is expected to be localized at the minimum of $\Delta E_\mathrm{tot}(\bm{r})$ corresponding to Site2 (the 4$c$ site).
The $z$ dependence of the simulated $B_{\rm loc}$ for the reported AFM structure is shown on the right axis of Fig.~\ref{Fig_Etot_jiba}. We show the two cases where the Ru moment $\lvert{\bm m}_\mathrm{Ru}\rvert=0.05\mu_\mathrm{B}$ with ${\bm m}_\mathrm{Ru}\parallel c$, and a 20$^\circ$ tilted to the (110) plane~\cite{Rxs_2019}. In the former case, $B_{\rm loc}$ cancels at $z=0.25$ (4$d$ site), but Mu is not likely to be stationary there because $\Delta E_\mathrm{tot}(\bm{r})$ is at a maximum.

Conversely, when the slow damping in the ZF spectrum at 5~K is interpreted as part of a precession signal associated with extremely small $B_{\rm loc}$ that appears with ordered Ru magnetic moments, we can estimate the maximum Ru moment size from the precession frequency $f$ obtained by curve fit using the following function,
\begin{align}
  G_z(t)=A_1\left[\tfrac{1}{3}+\tfrac{2}{3}\cos(2\pi ft+\varphi)\right],
  \label{Eq:ana_cos}
\end{align}
where $f=(\gamma_\mu/2\pi) B_\mathrm{loc}$, $\varphi$ (=0 in the present experimental setup) is the initial phase of the precession, $\gamma_\mu/2\pi=135.539$~MHz/T is the $\mu^+$ gyromagnetic ratio. Curve fit of the ZF spectrum at 5~K shown in Fig.~\ref{fig:spectra}(a) yields $f=8.0(2)\times10^{-3}$~MHz, corresponding to $B_\mathrm{loc}=5.9(2)\times10^{-2}$~mT. This value gives the upper boundary of $B_{\rm loc}$ for the field exerted from Ru $4d$ electrons. By comparison with the lowest value of $B_\mathrm{loc}$ among those obtained by the abovementioned simulation (5.96~mT at Site2), the upper limit for the $\lvert{\bm m}_\mathrm{Ru}\rvert$ can be estimated to be $4.8(2)\times10^{-4}\mu_\mathrm{B}$ for the reported magnetic structures. Following the same procedure, the upper limit for $\lvert{\bm m}_\mathrm{Ru}\rvert$ in the powder sample is also estimated to be $6.1(3)\times10^{-3}\mu_\mathrm{B}$ (see~\cite{sm} for details). 
Value for single crystal is the same order of magnitude as the nuclear magneton and is nearly two orders of magnitude smaller than the value of 0.05$\mu_\mathrm{B}$ reported in the earlier study~\cite{ND_2017}, strongly disfavoring the presumed magnetic order in the bulk crystal. It should be noted that these upper limits are even lower when the relaxation due to the {\it nuclear} magnetic moments of Ru ($\approx0.027$~$\mu$s$^{-1}$ at the Site1) is taken into account.

Finally, we briefly discuss a possibility of the slow relaxation due to the fast fluctuating magnetic order. The fluctuation time $\tau_c$ can be estimated using the Bloembergen-Purcell-Pound relation~\cite{BPP}, from the $\lambda_0\sim0.007$~$\mu$s$^{-1}$ in Eq.~(\ref{Eq:lmd_ana}) and $B_{\rm loc}\sim6$~mT,
  \begin{align*}
    \lambda_0(B_\mathrm{LF})\simeq2(\gamma_\mu B_\mathrm{loc})^2\tau_c/\left(1+(\gamma_\mu B_\mathrm{LF}\tau_c)^2\right),
  \end{align*}
yielding $\tau_c\sim0.13$~ns in ZF. While it appears static in neutron scattering, it is detectable in $\mu$SR. However, $B_\mathrm{LF}$ required to suppress the relaxation [$\lambda_0(B_\mathrm{LF})<0.001$~$\mu$s$^{-1}$] is estimated to be $>$20~T, which contradicts the experimental results, as shown in Fig.~\ref{fig:spectra}(b). Thus, the possibility of the magnetic order with fast fluctuation can be ruled out.

In summary, using $\mu$SR combined with DFT calculations, we provided the experimental evidence that the electronic ground state of bulk crystal RuO$_2$ is a nonmagnetic metal.
The $\mu$SR time spectrum at zero field shows only monotonous relaxation without sinusoidal oscillation, indicating that no homogeneous internal magnetic field is established at the Mu site(s). Our DFT calculations ruled out the possibility for Mu to occupy sites where the internal magnetic field accidentally cancels out for the reported AFM structure. The upper limit of the Ru moment size is nearly two orders of magnitude smaller than the reported value of 0.05$\mu_\mathrm{B}$. From these results, we conclude that there is no bulk AFM order as reported in RuO$_2$.

\begin{acknowledgments}
This work was supported by the MEXT Elements Strategy Initiative to Form Core Research Center for Electron Materials (Grant No. JPMXP0112101001) and JSPS KAKENHI (Grant No. 22K05275), and partially by the MEXT Program: Data Creation and Utilization Type Material Research and Development Project under Grant No. JPMXP1122683430. The authors would like to thank the help of the MUSE staff during $\mu$SR experiment at MLF, J-PARC. The experiment was conducted under the support of Inter-University Research Programs by Institute of Materials Structure Science, KEK (Proposal No. 2019MS02).
\end{acknowledgments}

\bibliography{Refs}

\begin{thebibliography}{49}%
\makeatletter
\providecommand \@ifxundefined [1]{%
 \@ifx{#1\undefined}
}%
\providecommand \@ifnum [1]{%
 \ifnum #1\expandafter \@firstoftwo
 \else \expandafter \@secondoftwo
 \fi
}%
\providecommand \@ifx [1]{%
 \ifx #1\expandafter \@firstoftwo
 \else \expandafter \@secondoftwo
 \fi
}%
\providecommand \natexlab [1]{#1}%
\providecommand \enquote  [1]{``#1''}%
\providecommand \bibnamefont  [1]{#1}%
\providecommand \bibfnamefont [1]{#1}%
\providecommand \citenamefont [1]{#1}%
\providecommand \href@noop [0]{\@secondoftwo}%
\providecommand \href [0]{\begingroup \@sanitize@url \@href}%
\providecommand \@href[1]{\@@startlink{#1}\@@href}%
\providecommand \@@href[1]{\endgroup#1\@@endlink}%
\providecommand \@sanitize@url [0]{\catcode `\\12\catcode `\$12\catcode
  `\&12\catcode `\#12\catcode `\^12\catcode `\_12\catcode `\%12\relax}%
\providecommand \@@startlink[1]{}%
\providecommand \@@endlink[0]{}%
\providecommand \url  [0]{\begingroup\@sanitize@url \@url }%
\providecommand \@url [1]{\endgroup\@href {#1}{\urlprefix }}%
\providecommand \urlprefix  [0]{URL }%
\providecommand \Eprint [0]{\href }%
\providecommand \doibase [0]{https://doi.org/}%
\providecommand \selectlanguage [0]{\@gobble}%
\providecommand \bibinfo  [0]{\@secondoftwo}%
\providecommand \bibfield  [0]{\@secondoftwo}%
\providecommand \translation [1]{[#1]}%
\providecommand \BibitemOpen [0]{}%
\providecommand \bibitemStop [0]{}%
\providecommand \bibitemNoStop [0]{.\EOS\space}%
\providecommand \EOS [0]{\spacefactor3000\relax}%
\providecommand \BibitemShut  [1]{\csname bibitem#1\endcsname}%
\let\auto@bib@innerbib\@empty
\bibitem [{\citenamefont {Over}(2012)}]{review2012}%
  \BibitemOpen
  \bibfield  {author} {\bibinfo {author} {\bibfnamefont {H.}~\bibnamefont
  {Over}},\ }\href {https://doi.org/10.1021/cr200247n} {\bibfield  {journal}
  {\bibinfo  {journal} {Chem. Rev.}\ }\textbf {\bibinfo {volume} {112}},\
  \bibinfo {pages} {3356} (\bibinfo {year} {2012})},\ \bibinfo {note} {pMID:
  22423981}\BibitemShut {NoStop}%
\bibitem [{\citenamefont {Ryden}\ and\ \citenamefont
  {Lawson}(2003)}]{Ryden_1970}%
  \BibitemOpen
  \bibfield  {author} {\bibinfo {author} {\bibfnamefont {W.~D.}\ \bibnamefont
  {Ryden}}\ and\ \bibinfo {author} {\bibfnamefont {A.~W.}\ \bibnamefont
  {Lawson}},\ }\href {https://doi.org/10.1063/1.1672908} {\bibfield  {journal}
  {\bibinfo  {journal} {J. Chem. Phys.}\ }\textbf {\bibinfo {volume} {52}},\
  \bibinfo {pages} {6058} (\bibinfo {year} {2003})}\BibitemShut {NoStop}%
\bibitem [{\citenamefont {Sun}\ \emph {et~al.}(2017)\citenamefont {Sun},
  \citenamefont {Zhang}, \citenamefont {Liu}, \citenamefont {Felser},\ and\
  \citenamefont {Yan}}]{Zhang2017}%
  \BibitemOpen
  \bibfield  {author} {\bibinfo {author} {\bibfnamefont {Y.}~\bibnamefont
  {Sun}}, \bibinfo {author} {\bibfnamefont {Y.}~\bibnamefont {Zhang}}, \bibinfo
  {author} {\bibfnamefont {C.-X.}\ \bibnamefont {Liu}}, \bibinfo {author}
  {\bibfnamefont {C.}~\bibnamefont {Felser}},\ and\ \bibinfo {author}
  {\bibfnamefont {B.}~\bibnamefont {Yan}},\ }\href
  {https://doi.org/10.1103/PhysRevB.95.235104} {\bibfield  {journal} {\bibinfo
  {journal} {Phys. Rev. B}\ }\textbf {\bibinfo {volume} {95}},\ \bibinfo
  {pages} {235104} (\bibinfo {year} {2017})}\BibitemShut {NoStop}%
\bibitem [{\citenamefont {Jovic}\ \emph {et~al.}(2018)\citenamefont {Jovic},
  \citenamefont {Koch}, \citenamefont {Panda}, \citenamefont {Berger},
  \citenamefont {Bugnon}, \citenamefont {Magrez}, \citenamefont {Smith},
  \citenamefont {Biermann}, \citenamefont {Jozwiak}, \citenamefont {Bostwick},
  \citenamefont {Rotenberg},\ and\ \citenamefont {Moser}}]{Vedran2018}%
  \BibitemOpen
  \bibfield  {author} {\bibinfo {author} {\bibfnamefont {V.}~\bibnamefont
  {Jovic}}, \bibinfo {author} {\bibfnamefont {R.~J.}\ \bibnamefont {Koch}},
  \bibinfo {author} {\bibfnamefont {S.~K.}\ \bibnamefont {Panda}}, \bibinfo
  {author} {\bibfnamefont {H.}~\bibnamefont {Berger}}, \bibinfo {author}
  {\bibfnamefont {P.}~\bibnamefont {Bugnon}}, \bibinfo {author} {\bibfnamefont
  {A.}~\bibnamefont {Magrez}}, \bibinfo {author} {\bibfnamefont {K.~E.}\
  \bibnamefont {Smith}}, \bibinfo {author} {\bibfnamefont {S.}~\bibnamefont
  {Biermann}}, \bibinfo {author} {\bibfnamefont {C.}~\bibnamefont {Jozwiak}},
  \bibinfo {author} {\bibfnamefont {A.}~\bibnamefont {Bostwick}}, \bibinfo
  {author} {\bibfnamefont {E.}~\bibnamefont {Rotenberg}},\ and\ \bibinfo
  {author} {\bibfnamefont {S.}~\bibnamefont {Moser}},\ }\href
  {https://doi.org/10.1103/PhysRevB.98.241101} {\bibfield  {journal} {\bibinfo
  {journal} {Phys. Rev. B}\ }\textbf {\bibinfo {volume} {98}},\ \bibinfo
  {pages} {241101(R)} (\bibinfo {year} {2018})}\BibitemShut {NoStop}%
\bibitem [{\citenamefont {Berlijn}\ \emph {et~al.}(2017)\citenamefont
  {Berlijn}, \citenamefont {Snijders}, \citenamefont {Delaire}, \citenamefont
  {Zhou}, \citenamefont {Maier}, \citenamefont {Cao}, \citenamefont {Chi},
  \citenamefont {Matsuda}, \citenamefont {Wang}, \citenamefont {Koehler},
  \citenamefont {Kent},\ and\ \citenamefont {Weitering}}]{ND_2017}%
  \BibitemOpen
  \bibfield  {author} {\bibinfo {author} {\bibfnamefont {T.}~\bibnamefont
  {Berlijn}}, \bibinfo {author} {\bibfnamefont {P.~C.}\ \bibnamefont
  {Snijders}}, \bibinfo {author} {\bibfnamefont {O.}~\bibnamefont {Delaire}},
  \bibinfo {author} {\bibfnamefont {H.-D.}\ \bibnamefont {Zhou}}, \bibinfo
  {author} {\bibfnamefont {T.~A.}\ \bibnamefont {Maier}}, \bibinfo {author}
  {\bibfnamefont {H.-B.}\ \bibnamefont {Cao}}, \bibinfo {author} {\bibfnamefont
  {S.-X.}\ \bibnamefont {Chi}}, \bibinfo {author} {\bibfnamefont
  {M.}~\bibnamefont {Matsuda}}, \bibinfo {author} {\bibfnamefont
  {Y.}~\bibnamefont {Wang}}, \bibinfo {author} {\bibfnamefont {M.~R.}\
  \bibnamefont {Koehler}}, \bibinfo {author} {\bibfnamefont {P.~R.~C.}\
  \bibnamefont {Kent}},\ and\ \bibinfo {author} {\bibfnamefont {H.~H.}\
  \bibnamefont {Weitering}},\ }\href
  {https://doi.org/10.1103/PhysRevLett.118.077201} {\bibfield  {journal}
  {\bibinfo  {journal} {Phys. Rev. Lett.}\ }\textbf {\bibinfo {volume} {118}},\
  \bibinfo {pages} {077201} (\bibinfo {year} {2017})}\BibitemShut {NoStop}%
\bibitem [{\citenamefont {Zhu}\ \emph {et~al.}(2019)\citenamefont {Zhu},
  \citenamefont {Strempfer}, \citenamefont {Rao}, \citenamefont {Occhialini},
  \citenamefont {Pelliciari}, \citenamefont {Choi}, \citenamefont {Kawaguchi},
  \citenamefont {You}, \citenamefont {Mitchell}, \citenamefont {Shao-Horn},\
  and\ \citenamefont {Comin}}]{Rxs_2019}%
  \BibitemOpen
  \bibfield  {author} {\bibinfo {author} {\bibfnamefont {Z.~H.}\ \bibnamefont
  {Zhu}}, \bibinfo {author} {\bibfnamefont {J.}~\bibnamefont {Strempfer}},
  \bibinfo {author} {\bibfnamefont {R.~R.}\ \bibnamefont {Rao}}, \bibinfo
  {author} {\bibfnamefont {C.~A.}\ \bibnamefont {Occhialini}}, \bibinfo
  {author} {\bibfnamefont {J.}~\bibnamefont {Pelliciari}}, \bibinfo {author}
  {\bibfnamefont {Y.}~\bibnamefont {Choi}}, \bibinfo {author} {\bibfnamefont
  {T.}~\bibnamefont {Kawaguchi}}, \bibinfo {author} {\bibfnamefont
  {H.}~\bibnamefont {You}}, \bibinfo {author} {\bibfnamefont {J.~F.}\
  \bibnamefont {Mitchell}}, \bibinfo {author} {\bibfnamefont {Y.}~\bibnamefont
  {Shao-Horn}},\ and\ \bibinfo {author} {\bibfnamefont {R.}~\bibnamefont
  {Comin}},\ }\href {https://doi.org/10.1103/PhysRevLett.122.017202} {\bibfield
   {journal} {\bibinfo  {journal} {Phys. Rev. Lett.}\ }\textbf {\bibinfo
  {volume} {122}},\ \bibinfo {pages} {017202} (\bibinfo {year}
  {2019})}\BibitemShut {NoStop}%
\bibitem [{\citenamefont {Wadley}\ \emph {et~al.}(2016)\citenamefont {Wadley},
  \citenamefont {Howells}, \citenamefont {\v{Z}elezn\'{y}}, \citenamefont
  {Andrews}, \citenamefont {Hills}, \citenamefont {Campion}, \citenamefont
  {Nov\'{a}k}, \citenamefont {Olejn\'{i}k}, \citenamefont {Maccherozzi},
  \citenamefont {Dhesi}, \citenamefont {Martin}, \citenamefont {Wagner},
  \citenamefont {Wunderlich}, \citenamefont {Freimuth}, \citenamefont
  {Mokrousov}, \citenamefont {Kune\v{s}}, \citenamefont {Chauhan},
  \citenamefont {Grzybowski}, \citenamefont {Rushforth}, \citenamefont
  {Edmonds}, \citenamefont {Gallagher},\ and\ \citenamefont
  {Jungwirth}}]{spintronics1}%
  \BibitemOpen
  \bibfield  {author} {\bibinfo {author} {\bibfnamefont {P.}~\bibnamefont
  {Wadley}}, \bibinfo {author} {\bibfnamefont {B.}~\bibnamefont {Howells}},
  \bibinfo {author} {\bibfnamefont {J.}~\bibnamefont {\v{Z}elezn\'{y}}},
  \bibinfo {author} {\bibfnamefont {C.}~\bibnamefont {Andrews}}, \bibinfo
  {author} {\bibfnamefont {V.}~\bibnamefont {Hills}}, \bibinfo {author}
  {\bibfnamefont {R.~P.}\ \bibnamefont {Campion}}, \bibinfo {author}
  {\bibfnamefont {V.}~\bibnamefont {Nov\'{a}k}}, \bibinfo {author}
  {\bibfnamefont {K.}~\bibnamefont {Olejn\'{i}k}}, \bibinfo {author}
  {\bibfnamefont {F.}~\bibnamefont {Maccherozzi}}, \bibinfo {author}
  {\bibfnamefont {S.~S.}\ \bibnamefont {Dhesi}}, \bibinfo {author}
  {\bibfnamefont {S.~Y.}\ \bibnamefont {Martin}}, \bibinfo {author}
  {\bibfnamefont {T.}~\bibnamefont {Wagner}}, \bibinfo {author} {\bibfnamefont
  {J.}~\bibnamefont {Wunderlich}}, \bibinfo {author} {\bibfnamefont
  {F.}~\bibnamefont {Freimuth}}, \bibinfo {author} {\bibfnamefont
  {Y.}~\bibnamefont {Mokrousov}}, \bibinfo {author} {\bibfnamefont
  {J.}~\bibnamefont {Kune\v{s}}}, \bibinfo {author} {\bibfnamefont {J.~S.}\
  \bibnamefont {Chauhan}}, \bibinfo {author} {\bibfnamefont {M.~J.}\
  \bibnamefont {Grzybowski}}, \bibinfo {author} {\bibfnamefont {A.~W.}\
  \bibnamefont {Rushforth}}, \bibinfo {author} {\bibfnamefont {K.~W.}\
  \bibnamefont {Edmonds}}, \bibinfo {author} {\bibfnamefont {B.~L.}\
  \bibnamefont {Gallagher}},\ and\ \bibinfo {author} {\bibfnamefont
  {T.}~\bibnamefont {Jungwirth}},\ }\href
  {https://doi.org/10.1126/science.aab1031} {\bibfield  {journal} {\bibinfo
  {journal} {Science}\ }\textbf {\bibinfo {volume} {351}},\ \bibinfo {pages}
  {587} (\bibinfo {year} {2016})}\BibitemShut {NoStop}%
\bibitem [{\citenamefont {Baltz}\ \emph {et~al.}(2018)\citenamefont {Baltz},
  \citenamefont {Manchon}, \citenamefont {Tsoi}, \citenamefont {Moriyama},
  \citenamefont {Ono},\ and\ \citenamefont {Tserkovnyak}}]{spintronics2}%
  \BibitemOpen
  \bibfield  {author} {\bibinfo {author} {\bibfnamefont {V.}~\bibnamefont
  {Baltz}}, \bibinfo {author} {\bibfnamefont {A.}~\bibnamefont {Manchon}},
  \bibinfo {author} {\bibfnamefont {M.}~\bibnamefont {Tsoi}}, \bibinfo {author}
  {\bibfnamefont {T.}~\bibnamefont {Moriyama}}, \bibinfo {author}
  {\bibfnamefont {T.}~\bibnamefont {Ono}},\ and\ \bibinfo {author}
  {\bibfnamefont {Y.}~\bibnamefont {Tserkovnyak}},\ }\href
  {https://doi.org/10.1103/RevModPhys.90.015005} {\bibfield  {journal}
  {\bibinfo  {journal} {Rev. Mod. Phys.}\ }\textbf {\bibinfo {volume} {90}},\
  \bibinfo {pages} {015005} (\bibinfo {year} {2018})}\BibitemShut {NoStop}%
\bibitem [{\citenamefont {Uchida}\ \emph {et~al.}(2020)\citenamefont {Uchida},
  \citenamefont {Nomoto}, \citenamefont {Musashi}, \citenamefont {Arita},\ and\
  \citenamefont {Kawasaki}}]{SC_2020}%
  \BibitemOpen
  \bibfield  {author} {\bibinfo {author} {\bibfnamefont {M.}~\bibnamefont
  {Uchida}}, \bibinfo {author} {\bibfnamefont {T.}~\bibnamefont {Nomoto}},
  \bibinfo {author} {\bibfnamefont {M.}~\bibnamefont {Musashi}}, \bibinfo
  {author} {\bibfnamefont {R.}~\bibnamefont {Arita}},\ and\ \bibinfo {author}
  {\bibfnamefont {M.}~\bibnamefont {Kawasaki}},\ }\href
  {https://doi.org/10.1103/PhysRevLett.125.147001} {\bibfield  {journal}
  {\bibinfo  {journal} {Phys. Rev. Lett.}\ }\textbf {\bibinfo {volume} {125}},\
  \bibinfo {pages} {147001} (\bibinfo {year} {2020})}\BibitemShut {NoStop}%
\bibitem [{\citenamefont {Ruf}\ \emph {et~al.}(2021)\citenamefont {Ruf},
  \citenamefont {Paik}, \citenamefont {Schreiber}, \citenamefont {Nair},
  \citenamefont {Miao}, \citenamefont {Kawasaki}, \citenamefont {Nelson},
  \citenamefont {Faeth}, \citenamefont {Lee}, \citenamefont {Goodge},
  \citenamefont {Pamuk}, \citenamefont {Fennie}, \citenamefont {Kourkoutis},
  \citenamefont {Schlom},\ and\ \citenamefont {Shen}}]{SC_2021}%
  \BibitemOpen
  \bibfield  {author} {\bibinfo {author} {\bibfnamefont {J.~P.}\ \bibnamefont
  {Ruf}}, \bibinfo {author} {\bibfnamefont {H.}~\bibnamefont {Paik}}, \bibinfo
  {author} {\bibfnamefont {N.~J.}\ \bibnamefont {Schreiber}}, \bibinfo {author}
  {\bibfnamefont {H.~P.}\ \bibnamefont {Nair}}, \bibinfo {author}
  {\bibfnamefont {L.}~\bibnamefont {Miao}}, \bibinfo {author} {\bibfnamefont
  {J.~K.}\ \bibnamefont {Kawasaki}}, \bibinfo {author} {\bibfnamefont {J.~N.}\
  \bibnamefont {Nelson}}, \bibinfo {author} {\bibfnamefont {B.~D.}\
  \bibnamefont {Faeth}}, \bibinfo {author} {\bibfnamefont {Y.}~\bibnamefont
  {Lee}}, \bibinfo {author} {\bibfnamefont {B.~H.}\ \bibnamefont {Goodge}},
  \bibinfo {author} {\bibfnamefont {B.}~\bibnamefont {Pamuk}}, \bibinfo
  {author} {\bibfnamefont {C.~J.}\ \bibnamefont {Fennie}}, \bibinfo {author}
  {\bibfnamefont {L.~F.}\ \bibnamefont {Kourkoutis}}, \bibinfo {author}
  {\bibfnamefont {D.~G.}\ \bibnamefont {Schlom}},\ and\ \bibinfo {author}
  {\bibfnamefont {K.~M.}\ \bibnamefont {Shen}},\ }\href
  {https://doi.org/10.1038/s41467-020-20252-7} {\bibfield  {journal} {\bibinfo
  {journal} {Nat. Commun.}\ }\textbf {\bibinfo {volume} {12}},\ \bibinfo
  {pages} {59} (\bibinfo {year} {2021})}\BibitemShut {NoStop}%
\bibitem [{\citenamefont {$\mathrm{\check{S}}$mejkal}\ \emph
  {et~al.}(2020)\citenamefont {$\mathrm{\check{S}}$mejkal}, \citenamefont
  {Gonz$\mathrm{\acute{a}}$lez-Hern$\mathrm{\acute{a}}$ndez}, \citenamefont
  {Jungwirth},\ and\ \citenamefont {Sinova}}]{Science_2020_AHE_theory}%
  \BibitemOpen
  \bibfield  {author} {\bibinfo {author} {\bibfnamefont {L.}~\bibnamefont
  {$\mathrm{\check{S}}$mejkal}}, \bibinfo {author} {\bibfnamefont
  {R.}~\bibnamefont
  {Gonz$\mathrm{\acute{a}}$lez-Hern$\mathrm{\acute{a}}$ndez}}, \bibinfo
  {author} {\bibfnamefont {T.}~\bibnamefont {Jungwirth}},\ and\ \bibinfo
  {author} {\bibfnamefont {J.}~\bibnamefont {Sinova}},\ }\href
  {https://doi.org/10.1126/sciadv.aaz8809} {\bibfield  {journal} {\bibinfo
  {journal} {Sci. Adv.}\ }\textbf {\bibinfo {volume} {6}},\ \bibinfo {pages}
  {eaaz8809} (\bibinfo {year} {2020})}\BibitemShut {NoStop}%
\bibitem [{\citenamefont {Feng}\ \emph {et~al.}(2022)\citenamefont {Feng},
  \citenamefont {Zhou}, \citenamefont {{\v{S}}mejkal}, \citenamefont {Wu},
  \citenamefont {Zhu}, \citenamefont {Guo}, \citenamefont
  {Gonz{\'a}lez-Hern{\'a}ndez}, \citenamefont {Wang}, \citenamefont {Yan},
  \citenamefont {Qin}, \citenamefont {Zhang}, \citenamefont {Wu}, \citenamefont
  {Chen}, \citenamefont {Meng}, \citenamefont {Liu}, \citenamefont {Xia},
  \citenamefont {Sinova}, \citenamefont {Jungwirth},\ and\ \citenamefont
  {Liu}}]{Feng2022_AHE}%
  \BibitemOpen
  \bibfield  {author} {\bibinfo {author} {\bibfnamefont {Z.}~\bibnamefont
  {Feng}}, \bibinfo {author} {\bibfnamefont {X.}~\bibnamefont {Zhou}}, \bibinfo
  {author} {\bibfnamefont {L.}~\bibnamefont {{\v{S}}mejkal}}, \bibinfo {author}
  {\bibfnamefont {L.}~\bibnamefont {Wu}}, \bibinfo {author} {\bibfnamefont
  {Z.}~\bibnamefont {Zhu}}, \bibinfo {author} {\bibfnamefont {H.}~\bibnamefont
  {Guo}}, \bibinfo {author} {\bibfnamefont {R.}~\bibnamefont
  {Gonz{\'a}lez-Hern{\'a}ndez}}, \bibinfo {author} {\bibfnamefont
  {X.}~\bibnamefont {Wang}}, \bibinfo {author} {\bibfnamefont {H.}~\bibnamefont
  {Yan}}, \bibinfo {author} {\bibfnamefont {P.}~\bibnamefont {Qin}}, \bibinfo
  {author} {\bibfnamefont {X.}~\bibnamefont {Zhang}}, \bibinfo {author}
  {\bibfnamefont {H.}~\bibnamefont {Wu}}, \bibinfo {author} {\bibfnamefont
  {H.}~\bibnamefont {Chen}}, \bibinfo {author} {\bibfnamefont {Z.}~\bibnamefont
  {Meng}}, \bibinfo {author} {\bibfnamefont {L.}~\bibnamefont {Liu}}, \bibinfo
  {author} {\bibfnamefont {Z.}~\bibnamefont {Xia}}, \bibinfo {author}
  {\bibfnamefont {J.}~\bibnamefont {Sinova}}, \bibinfo {author} {\bibfnamefont
  {T.}~\bibnamefont {Jungwirth}},\ and\ \bibinfo {author} {\bibfnamefont
  {Z.}~\bibnamefont {Liu}},\ }\href
  {https://doi.org/10.1038/s41928-022-00866-z} {\bibfield  {journal} {\bibinfo
  {journal} {Nat. Electron.}\ }\textbf {\bibinfo {volume} {5}},\ \bibinfo
  {pages} {735} (\bibinfo {year} {2022})}\BibitemShut {NoStop}%
\bibitem [{\citenamefont {Gonz\'alez-Hern\'andez}\ \emph
  {et~al.}(2021)\citenamefont {Gonz\'alez-Hern\'andez}, \citenamefont
  {\ifmmode~\check{S}\else \v{S}\fi{}mejkal}, \citenamefont {V\'yborn\'y},
  \citenamefont {Yahagi}, \citenamefont {Sinova}, \citenamefont {Jungwirth},\
  and\ \citenamefont {\ifmmode~\check{Z}\else
  \v{Z}\fi{}elezn\'y}}]{SST_theory_2021}%
  \BibitemOpen
  \bibfield  {author} {\bibinfo {author} {\bibfnamefont {R.}~\bibnamefont
  {Gonz\'alez-Hern\'andez}}, \bibinfo {author} {\bibfnamefont {L.}~\bibnamefont
  {\ifmmode~\check{S}\else \v{S}\fi{}mejkal}}, \bibinfo {author} {\bibfnamefont
  {K.}~\bibnamefont {V\'yborn\'y}}, \bibinfo {author} {\bibfnamefont
  {Y.}~\bibnamefont {Yahagi}}, \bibinfo {author} {\bibfnamefont
  {J.}~\bibnamefont {Sinova}}, \bibinfo {author} {\bibfnamefont {T.~c.~v.}\
  \bibnamefont {Jungwirth}},\ and\ \bibinfo {author} {\bibfnamefont
  {J.}~\bibnamefont {\ifmmode~\check{Z}\else \v{Z}\fi{}elezn\'y}},\ }\href
  {https://doi.org/10.1103/PhysRevLett.126.127701} {\bibfield  {journal}
  {\bibinfo  {journal} {Phys. Rev. Lett.}\ }\textbf {\bibinfo {volume} {126}},\
  \bibinfo {pages} {127701} (\bibinfo {year} {2021})}\BibitemShut {NoStop}%
\bibitem [{\citenamefont {Bai}\ \emph {et~al.}(2022)\citenamefont {Bai},
  \citenamefont {Han}, \citenamefont {Feng}, \citenamefont {Zhou},
  \citenamefont {Su}, \citenamefont {Wang}, \citenamefont {Liao}, \citenamefont
  {Zhu}, \citenamefont {Chen}, \citenamefont {Pan}, \citenamefont {Fan},\ and\
  \citenamefont {Song}}]{SST_exp2_2022}%
  \BibitemOpen
  \bibfield  {author} {\bibinfo {author} {\bibfnamefont {H.}~\bibnamefont
  {Bai}}, \bibinfo {author} {\bibfnamefont {L.}~\bibnamefont {Han}}, \bibinfo
  {author} {\bibfnamefont {X.~Y.}\ \bibnamefont {Feng}}, \bibinfo {author}
  {\bibfnamefont {Y.~J.}\ \bibnamefont {Zhou}}, \bibinfo {author}
  {\bibfnamefont {R.~X.}\ \bibnamefont {Su}}, \bibinfo {author} {\bibfnamefont
  {Q.}~\bibnamefont {Wang}}, \bibinfo {author} {\bibfnamefont {L.~Y.}\
  \bibnamefont {Liao}}, \bibinfo {author} {\bibfnamefont {W.~X.}\ \bibnamefont
  {Zhu}}, \bibinfo {author} {\bibfnamefont {X.~Z.}\ \bibnamefont {Chen}},
  \bibinfo {author} {\bibfnamefont {F.}~\bibnamefont {Pan}}, \bibinfo {author}
  {\bibfnamefont {X.~L.}\ \bibnamefont {Fan}},\ and\ \bibinfo {author}
  {\bibfnamefont {C.}~\bibnamefont {Song}},\ }\href
  {https://doi.org/10.1103/PhysRevLett.128.197202} {\bibfield  {journal}
  {\bibinfo  {journal} {Phys. Rev. Lett.}\ }\textbf {\bibinfo {volume} {128}},\
  \bibinfo {pages} {197202} (\bibinfo {year} {2022})}\BibitemShut {NoStop}%
\bibitem [{\citenamefont {Bose}\ \emph {et~al.}(2022)\citenamefont {Bose},
  \citenamefont {Schreiber}, \citenamefont {Jain}, \citenamefont {Shao},
  \citenamefont {Nair}, \citenamefont {Sun}, \citenamefont {Zhang},
  \citenamefont {Muller}, \citenamefont {Tsymbal}, \citenamefont {Schlom},\
  and\ \citenamefont {Ralph}}]{SST_exp3_2022}%
  \BibitemOpen
  \bibfield  {author} {\bibinfo {author} {\bibfnamefont {A.}~\bibnamefont
  {Bose}}, \bibinfo {author} {\bibfnamefont {N.~J.}\ \bibnamefont {Schreiber}},
  \bibinfo {author} {\bibfnamefont {R.}~\bibnamefont {Jain}}, \bibinfo {author}
  {\bibfnamefont {D.-F.}\ \bibnamefont {Shao}}, \bibinfo {author}
  {\bibfnamefont {H.~P.}\ \bibnamefont {Nair}}, \bibinfo {author}
  {\bibfnamefont {J.}~\bibnamefont {Sun}}, \bibinfo {author} {\bibfnamefont
  {X.~S.}\ \bibnamefont {Zhang}}, \bibinfo {author} {\bibfnamefont {D.~A.}\
  \bibnamefont {Muller}}, \bibinfo {author} {\bibfnamefont {E.~Y.}\
  \bibnamefont {Tsymbal}}, \bibinfo {author} {\bibfnamefont {D.~G.}\
  \bibnamefont {Schlom}},\ and\ \bibinfo {author} {\bibfnamefont {D.~C.}\
  \bibnamefont {Ralph}},\ }\href {https://doi.org/10.1038/s41928-022-00744-8}
  {\bibfield  {journal} {\bibinfo  {journal} {Nat. Electron.}\ }\textbf
  {\bibinfo {volume} {5}},\ \bibinfo {pages} {267} (\bibinfo {year}
  {2022})}\BibitemShut {NoStop}%
\bibitem [{\citenamefont {Karube}\ \emph {et~al.}(2022)\citenamefont {Karube},
  \citenamefont {Tanaka}, \citenamefont {Sugawara}, \citenamefont {Kadoguchi},
  \citenamefont {Kohda},\ and\ \citenamefont {Nitta}}]{SST_exp1_2022}%
  \BibitemOpen
  \bibfield  {author} {\bibinfo {author} {\bibfnamefont {S.}~\bibnamefont
  {Karube}}, \bibinfo {author} {\bibfnamefont {T.}~\bibnamefont {Tanaka}},
  \bibinfo {author} {\bibfnamefont {D.}~\bibnamefont {Sugawara}}, \bibinfo
  {author} {\bibfnamefont {N.}~\bibnamefont {Kadoguchi}}, \bibinfo {author}
  {\bibfnamefont {M.}~\bibnamefont {Kohda}},\ and\ \bibinfo {author}
  {\bibfnamefont {J.}~\bibnamefont {Nitta}},\ }\href
  {https://doi.org/10.1103/PhysRevLett.129.137201} {\bibfield  {journal}
  {\bibinfo  {journal} {Phys. Rev. Lett.}\ }\textbf {\bibinfo {volume} {129}},\
  \bibinfo {pages} {137201} (\bibinfo {year} {2022})}\BibitemShut {NoStop}%
\bibitem [{\citenamefont {Zhou}\ \emph {et~al.}(2021)\citenamefont {Zhou},
  \citenamefont {Feng}, \citenamefont {Yang}, \citenamefont {Guo},\ and\
  \citenamefont {Yao}}]{CCMO_21}%
  \BibitemOpen
  \bibfield  {author} {\bibinfo {author} {\bibfnamefont {X.}~\bibnamefont
  {Zhou}}, \bibinfo {author} {\bibfnamefont {W.}~\bibnamefont {Feng}}, \bibinfo
  {author} {\bibfnamefont {X.}~\bibnamefont {Yang}}, \bibinfo {author}
  {\bibfnamefont {G.-Y.}\ \bibnamefont {Guo}},\ and\ \bibinfo {author}
  {\bibfnamefont {Y.}~\bibnamefont {Yao}},\ }\href
  {https://doi.org/10.1103/PhysRevB.104.024401} {\bibfield  {journal} {\bibinfo
   {journal} {Phys. Rev. B}\ }\textbf {\bibinfo {volume} {104}},\ \bibinfo
  {pages} {024401} (\bibinfo {year} {2021})}\BibitemShut {NoStop}%
\bibitem [{\citenamefont {Smolyanyuk}\ \emph {et~al.}(2023)\citenamefont
  {Smolyanyuk}, \citenamefont {Mazin}, \citenamefont {Garcia-Gassull},\ and\
  \citenamefont {Valent^^c3^^ad}}]{arxiv_mazin_23}%
  \BibitemOpen
  \bibfield  {author} {\bibinfo {author} {\bibfnamefont {A.}~\bibnamefont
  {Smolyanyuk}}, \bibinfo {author} {\bibfnamefont {I.~I.}\ \bibnamefont
  {Mazin}}, \bibinfo {author} {\bibfnamefont {L.}~\bibnamefont
  {Garcia-Gassull}},\ and\ \bibinfo {author} {\bibfnamefont {R.}~\bibnamefont
  {Valent^^c3^^ad}},\ }\href@noop {} {\bibfield  {journal} {\bibinfo  {journal}
  {arXiv}\ } (\bibinfo {year} {2023})},\ \Eprint
  {https://arxiv.org/abs/2310.06909} {arXiv:2310.06909} \BibitemShut {NoStop}%
\bibitem [{\citenamefont {Nishida}\ \emph {et~al.}(1987)\citenamefont
  {Nishida}, \citenamefont {Miyatake}, \citenamefont {Shimada}, \citenamefont
  {Okuma}, \citenamefont {Ishikawa}, \citenamefont {Takabatake}, \citenamefont
  {Nakazawa}, \citenamefont {Kuno}, \citenamefont {Keitel}, \citenamefont
  {Brewer}, \citenamefont {Riseman}, \citenamefont {Williams}, \citenamefont
  {Watanabe}, \citenamefont {Yamazaki}, \citenamefont {Nishiyama},
  \citenamefont {Nagamine}, \citenamefont {Ansaldo},\ and\ \citenamefont
  {Torikai}}]{Nishida_1987}%
  \BibitemOpen
  \bibfield  {author} {\bibinfo {author} {\bibfnamefont {N.}~\bibnamefont
  {Nishida}}, \bibinfo {author} {\bibfnamefont {H.}~\bibnamefont {Miyatake}},
  \bibinfo {author} {\bibfnamefont {D.}~\bibnamefont {Shimada}}, \bibinfo
  {author} {\bibfnamefont {S.}~\bibnamefont {Okuma}}, \bibinfo {author}
  {\bibfnamefont {M.}~\bibnamefont {Ishikawa}}, \bibinfo {author}
  {\bibfnamefont {T.}~\bibnamefont {Takabatake}}, \bibinfo {author}
  {\bibfnamefont {Y.}~\bibnamefont {Nakazawa}}, \bibinfo {author}
  {\bibfnamefont {Y.}~\bibnamefont {Kuno}}, \bibinfo {author} {\bibfnamefont
  {R.}~\bibnamefont {Keitel}}, \bibinfo {author} {\bibfnamefont {J.~H.}\
  \bibnamefont {Brewer}}, \bibinfo {author} {\bibfnamefont {T.~M.}\
  \bibnamefont {Riseman}}, \bibinfo {author} {\bibfnamefont {D.~L.}\
  \bibnamefont {Williams}}, \bibinfo {author} {\bibfnamefont {Y.}~\bibnamefont
  {Watanabe}}, \bibinfo {author} {\bibfnamefont {T.}~\bibnamefont {Yamazaki}},
  \bibinfo {author} {\bibfnamefont {K.}~\bibnamefont {Nishiyama}}, \bibinfo
  {author} {\bibfnamefont {K.}~\bibnamefont {Nagamine}}, \bibinfo {author}
  {\bibfnamefont {E.~J.}\ \bibnamefont {Ansaldo}},\ and\ \bibinfo {author}
  {\bibfnamefont {E.}~\bibnamefont {Torikai}},\ }\href
  {https://doi.org/10.1143/JJAP.26.L1856} {\bibfield  {journal} {\bibinfo
  {journal} {Jpn. J. Appl. Phys.}\ }\textbf {\bibinfo {volume} {26}},\ \bibinfo
  {pages} {L1856} (\bibinfo {year} {1987})}\BibitemShut {NoStop}%
\bibitem [{\citenamefont {Uemura}\ \emph {et~al.}(1987)\citenamefont {Uemura},
  \citenamefont {Kossler}, \citenamefont {Yu}, \citenamefont {Kempton},
  \citenamefont {Schone}, \citenamefont {Opie}, \citenamefont {Stronach},
  \citenamefont {Johnston}, \citenamefont {Alvarez},\ and\ \citenamefont
  {Goshorn}}]{Uemura_1987}%
  \BibitemOpen
  \bibfield  {author} {\bibinfo {author} {\bibfnamefont {Y.~J.}\ \bibnamefont
  {Uemura}}, \bibinfo {author} {\bibfnamefont {W.~J.}\ \bibnamefont {Kossler}},
  \bibinfo {author} {\bibfnamefont {X.~H.}\ \bibnamefont {Yu}}, \bibinfo
  {author} {\bibfnamefont {J.~R.}\ \bibnamefont {Kempton}}, \bibinfo {author}
  {\bibfnamefont {H.~E.}\ \bibnamefont {Schone}}, \bibinfo {author}
  {\bibfnamefont {D.}~\bibnamefont {Opie}}, \bibinfo {author} {\bibfnamefont
  {C.~E.}\ \bibnamefont {Stronach}}, \bibinfo {author} {\bibfnamefont {D.~C.}\
  \bibnamefont {Johnston}}, \bibinfo {author} {\bibfnamefont {M.~S.}\
  \bibnamefont {Alvarez}},\ and\ \bibinfo {author} {\bibfnamefont {D.~P.}\
  \bibnamefont {Goshorn}},\ }\href
  {https://doi.org/10.1103/PhysRevLett.59.1045} {\bibfield  {journal} {\bibinfo
   {journal} {Phys. Rev. Lett.}\ }\textbf {\bibinfo {volume} {59}},\ \bibinfo
  {pages} {1045} (\bibinfo {year} {1987})}\BibitemShut {NoStop}%
\bibitem [{\citenamefont {Luke}\ \emph {et~al.}(1989)\citenamefont {Luke},
  \citenamefont {Sternlieb}, \citenamefont {Uemura}, \citenamefont {Brewer},
  \citenamefont {Kadono}, \citenamefont {Kiefl}, \citenamefont {Kreitzman},
  \citenamefont {Riseman}, \citenamefont {Gopalakrishnan}, \citenamefont
  {Sleight}, \citenamefont {Subramanian}, \citenamefont {Uchida}, \citenamefont
  {Takagi},\ and\ \citenamefont {Tokura}}]{Luke:89}%
  \BibitemOpen
  \bibfield  {author} {\bibinfo {author} {\bibfnamefont {G.~M.}\ \bibnamefont
  {Luke}}, \bibinfo {author} {\bibfnamefont {B.~J.}\ \bibnamefont {Sternlieb}},
  \bibinfo {author} {\bibfnamefont {Y.~J.}\ \bibnamefont {Uemura}}, \bibinfo
  {author} {\bibfnamefont {J.~H.}\ \bibnamefont {Brewer}}, \bibinfo {author}
  {\bibfnamefont {R.}~\bibnamefont {Kadono}}, \bibinfo {author} {\bibfnamefont
  {R.~F.}\ \bibnamefont {Kiefl}}, \bibinfo {author} {\bibfnamefont {S.~R.}\
  \bibnamefont {Kreitzman}}, \bibinfo {author} {\bibfnamefont {T.~M.}\
  \bibnamefont {Riseman}}, \bibinfo {author} {\bibfnamefont {J.}~\bibnamefont
  {Gopalakrishnan}}, \bibinfo {author} {\bibfnamefont {A.~W.}\ \bibnamefont
  {Sleight}}, \bibinfo {author} {\bibfnamefont {M.~A.}\ \bibnamefont
  {Subramanian}}, \bibinfo {author} {\bibfnamefont {S.}~\bibnamefont {Uchida}},
  \bibinfo {author} {\bibfnamefont {H.}~\bibnamefont {Takagi}},\ and\ \bibinfo
  {author} {\bibfnamefont {Y.}~\bibnamefont {Tokura}},\ }\href@noop {}
  {\bibfield  {journal} {\bibinfo  {journal} {Nature}\ }\textbf {\bibinfo
  {volume} {338}},\ \bibinfo {pages} {49} (\bibinfo {year} {1989})}\BibitemShut
  {NoStop}%
\bibitem [{\citenamefont {Aeppli}\ \emph {et~al.}(1987)\citenamefont {Aeppli},
  \citenamefont {Cava}, \citenamefont {Ansaldo}, \citenamefont {Brewer},
  \citenamefont {Kreitzman}, \citenamefont {Luke}, \citenamefont {Noakes},\
  and\ \citenamefont {Kiefl}}]{Appeli:87}%
  \BibitemOpen
  \bibfield  {author} {\bibinfo {author} {\bibfnamefont {G.}~\bibnamefont
  {Aeppli}}, \bibinfo {author} {\bibfnamefont {R.~J.}\ \bibnamefont {Cava}},
  \bibinfo {author} {\bibfnamefont {E.~J.}\ \bibnamefont {Ansaldo}}, \bibinfo
  {author} {\bibfnamefont {J.~H.}\ \bibnamefont {Brewer}}, \bibinfo {author}
  {\bibfnamefont {S.~R.}\ \bibnamefont {Kreitzman}}, \bibinfo {author}
  {\bibfnamefont {G.~M.}\ \bibnamefont {Luke}}, \bibinfo {author}
  {\bibfnamefont {D.~R.}\ \bibnamefont {Noakes}},\ and\ \bibinfo {author}
  {\bibfnamefont {R.~F.}\ \bibnamefont {Kiefl}},\ }\href
  {https://doi.org/10.1103/PhysRevB.35.7129} {\bibfield  {journal} {\bibinfo
  {journal} {Phys. Rev. B}\ }\textbf {\bibinfo {volume} {35}},\ \bibinfo
  {pages} {7129} (\bibinfo {year} {1987})}\BibitemShut {NoStop}%
\bibitem [{\citenamefont {Uemura}\ \emph {et~al.}(1989)\citenamefont {Uemura},
  \citenamefont {Luke}, \citenamefont {Sternlieb}, \citenamefont {Brewer},
  \citenamefont {Carolan}, \citenamefont {Hardy}, \citenamefont {Kadono},
  \citenamefont {Kempton}, \citenamefont {Kiefl}, \citenamefont {Kreitzman},
  \citenamefont {Mulhern}, \citenamefont {Riseman}, \citenamefont {Williams},
  \citenamefont {Yang}, \citenamefont {Uchida}, \citenamefont {Takagi},
  \citenamefont {Gopalakrishnan}, \citenamefont {Sleight}, \citenamefont
  {Subramanian}, \citenamefont {Chien}, \citenamefont {Cieplak}, \citenamefont
  {Xiao}, \citenamefont {Lee}, \citenamefont {Statt}, \citenamefont {Stronach},
  \citenamefont {Kossler},\ and\ \citenamefont {Yu}}]{Uemura:89}%
  \BibitemOpen
  \bibfield  {author} {\bibinfo {author} {\bibfnamefont {Y.~J.}\ \bibnamefont
  {Uemura}}, \bibinfo {author} {\bibfnamefont {G.~M.}\ \bibnamefont {Luke}},
  \bibinfo {author} {\bibfnamefont {B.~J.}\ \bibnamefont {Sternlieb}}, \bibinfo
  {author} {\bibfnamefont {J.~H.}\ \bibnamefont {Brewer}}, \bibinfo {author}
  {\bibfnamefont {J.~F.}\ \bibnamefont {Carolan}}, \bibinfo {author}
  {\bibfnamefont {W.~N.}\ \bibnamefont {Hardy}}, \bibinfo {author}
  {\bibfnamefont {R.}~\bibnamefont {Kadono}}, \bibinfo {author} {\bibfnamefont
  {J.~R.}\ \bibnamefont {Kempton}}, \bibinfo {author} {\bibfnamefont {R.~F.}\
  \bibnamefont {Kiefl}}, \bibinfo {author} {\bibfnamefont {S.~R.}\ \bibnamefont
  {Kreitzman}}, \bibinfo {author} {\bibfnamefont {P.}~\bibnamefont {Mulhern}},
  \bibinfo {author} {\bibfnamefont {T.~M.}\ \bibnamefont {Riseman}}, \bibinfo
  {author} {\bibfnamefont {D.~L.}\ \bibnamefont {Williams}}, \bibinfo {author}
  {\bibfnamefont {B.~X.}\ \bibnamefont {Yang}}, \bibinfo {author}
  {\bibfnamefont {S.}~\bibnamefont {Uchida}}, \bibinfo {author} {\bibfnamefont
  {H.}~\bibnamefont {Takagi}}, \bibinfo {author} {\bibfnamefont
  {J.}~\bibnamefont {Gopalakrishnan}}, \bibinfo {author} {\bibfnamefont
  {A.~W.}\ \bibnamefont {Sleight}}, \bibinfo {author} {\bibfnamefont {M.~A.}\
  \bibnamefont {Subramanian}}, \bibinfo {author} {\bibfnamefont {C.~L.}\
  \bibnamefont {Chien}}, \bibinfo {author} {\bibfnamefont {M.~Z.}\ \bibnamefont
  {Cieplak}}, \bibinfo {author} {\bibfnamefont {G.}~\bibnamefont {Xiao}},
  \bibinfo {author} {\bibfnamefont {V.~Y.}\ \bibnamefont {Lee}}, \bibinfo
  {author} {\bibfnamefont {B.~W.}\ \bibnamefont {Statt}}, \bibinfo {author}
  {\bibfnamefont {C.~E.}\ \bibnamefont {Stronach}}, \bibinfo {author}
  {\bibfnamefont {W.~J.}\ \bibnamefont {Kossler}},\ and\ \bibinfo {author}
  {\bibfnamefont {X.~H.}\ \bibnamefont {Yu}},\ }\href
  {https://doi.org/10.1103/PhysRevLett.62.2317} {\bibfield  {journal} {\bibinfo
   {journal} {Phys. Rev. Lett.}\ }\textbf {\bibinfo {volume} {62}},\ \bibinfo
  {pages} {2317} (\bibinfo {year} {1989})}\BibitemShut {NoStop}%
\bibitem [{\citenamefont {Sonier}\ \emph {et~al.}(2000)\citenamefont {Sonier},
  \citenamefont {Brewer},\ and\ \citenamefont {Kiefl}}]{Sonier:00}%
  \BibitemOpen
  \bibfield  {author} {\bibinfo {author} {\bibfnamefont {J.~E.}\ \bibnamefont
  {Sonier}}, \bibinfo {author} {\bibfnamefont {J.~H.}\ \bibnamefont {Brewer}},\
  and\ \bibinfo {author} {\bibfnamefont {R.~F.}\ \bibnamefont {Kiefl}},\ }\href
  {https://doi.org/10.1103/RevModPhys.72.769} {\bibfield  {journal} {\bibinfo
  {journal} {Rev. Mod. Phys.}\ }\textbf {\bibinfo {volume} {72}},\ \bibinfo
  {pages} {769} (\bibinfo {year} {2000})}\BibitemShut {NoStop}%
\bibitem [{\citenamefont {Giannozzi}\ \emph {et~al.}(2009)\citenamefont
  {Giannozzi}, \citenamefont {Baroni}, \citenamefont {Bonini}, \citenamefont
  {Calandra}, \citenamefont {Car}, \citenamefont {Cavazzoni}, \citenamefont
  {Ceresoli}, \citenamefont {Chiarotti}, \citenamefont {Cococcioni},
  \citenamefont {Dabo}, \citenamefont {Corso}, \citenamefont {de~Gironcoli},
  \citenamefont {Fabris}, \citenamefont {Fratesi}, \citenamefont {Gebauer},
  \citenamefont {Gerstmann}, \citenamefont {Gougoussis}, \citenamefont
  {Kokalj}, \citenamefont {Lazzeri}, \citenamefont {Martin-Samos},
  \citenamefont {Marzari}, \citenamefont {Mauri}, \citenamefont {Mazzarello},
  \citenamefont {Paolini}, \citenamefont {Pasquarello}, \citenamefont
  {Paulatto}, \citenamefont {Sbraccia}, \citenamefont {Scandolo}, \citenamefont
  {Sclauzero}, \citenamefont {Seitsonen}, \citenamefont {Smogunov},
  \citenamefont {Umari},\ and\ \citenamefont {Wentzcovitch}}]{QE_2009}%
  \BibitemOpen
  \bibfield  {author} {\bibinfo {author} {\bibfnamefont {P.}~\bibnamefont
  {Giannozzi}}, \bibinfo {author} {\bibfnamefont {S.}~\bibnamefont {Baroni}},
  \bibinfo {author} {\bibfnamefont {N.}~\bibnamefont {Bonini}}, \bibinfo
  {author} {\bibfnamefont {M.}~\bibnamefont {Calandra}}, \bibinfo {author}
  {\bibfnamefont {R.}~\bibnamefont {Car}}, \bibinfo {author} {\bibfnamefont
  {C.}~\bibnamefont {Cavazzoni}}, \bibinfo {author} {\bibfnamefont
  {D.}~\bibnamefont {Ceresoli}}, \bibinfo {author} {\bibfnamefont {G.~L.}\
  \bibnamefont {Chiarotti}}, \bibinfo {author} {\bibfnamefont {M.}~\bibnamefont
  {Cococcioni}}, \bibinfo {author} {\bibfnamefont {I.}~\bibnamefont {Dabo}},
  \bibinfo {author} {\bibfnamefont {A.~D.}\ \bibnamefont {Corso}}, \bibinfo
  {author} {\bibfnamefont {S.}~\bibnamefont {de~Gironcoli}}, \bibinfo {author}
  {\bibfnamefont {S.}~\bibnamefont {Fabris}}, \bibinfo {author} {\bibfnamefont
  {G.}~\bibnamefont {Fratesi}}, \bibinfo {author} {\bibfnamefont
  {R.}~\bibnamefont {Gebauer}}, \bibinfo {author} {\bibfnamefont
  {U.}~\bibnamefont {Gerstmann}}, \bibinfo {author} {\bibfnamefont
  {C.}~\bibnamefont {Gougoussis}}, \bibinfo {author} {\bibfnamefont
  {A.}~\bibnamefont {Kokalj}}, \bibinfo {author} {\bibfnamefont
  {M.}~\bibnamefont {Lazzeri}}, \bibinfo {author} {\bibfnamefont
  {L.}~\bibnamefont {Martin-Samos}}, \bibinfo {author} {\bibfnamefont
  {N.}~\bibnamefont {Marzari}}, \bibinfo {author} {\bibfnamefont
  {F.}~\bibnamefont {Mauri}}, \bibinfo {author} {\bibfnamefont
  {R.}~\bibnamefont {Mazzarello}}, \bibinfo {author} {\bibfnamefont
  {S.}~\bibnamefont {Paolini}}, \bibinfo {author} {\bibfnamefont
  {A.}~\bibnamefont {Pasquarello}}, \bibinfo {author} {\bibfnamefont
  {L.}~\bibnamefont {Paulatto}}, \bibinfo {author} {\bibfnamefont
  {C.}~\bibnamefont {Sbraccia}}, \bibinfo {author} {\bibfnamefont
  {S.}~\bibnamefont {Scandolo}}, \bibinfo {author} {\bibfnamefont
  {G.}~\bibnamefont {Sclauzero}}, \bibinfo {author} {\bibfnamefont {A.~P.}\
  \bibnamefont {Seitsonen}}, \bibinfo {author} {\bibfnamefont {A.}~\bibnamefont
  {Smogunov}}, \bibinfo {author} {\bibfnamefont {P.}~\bibnamefont {Umari}},\
  and\ \bibinfo {author} {\bibfnamefont {R.~M.}\ \bibnamefont {Wentzcovitch}},\
  }\href {https://doi.org/10.1088/0953-8984/21/39/395502} {\bibfield  {journal}
  {\bibinfo  {journal} {J. Phys.: Condens. Matter}\ }\textbf {\bibinfo {volume}
  {21}},\ \bibinfo {pages} {395502} (\bibinfo {year} {2009})}\BibitemShut
  {NoStop}%
\bibitem [{\citenamefont {Giannozzi}\ \emph {et~al.}(2017)\citenamefont
  {Giannozzi}, \citenamefont {Andreussi}, \citenamefont {Brumme}, \citenamefont
  {Bunau}, \citenamefont {Nardelli}, \citenamefont {Calandra}, \citenamefont
  {Car}, \citenamefont {Cavazzoni}, \citenamefont {Ceresoli}, \citenamefont
  {Cococcioni}, \citenamefont {Colonna}, \citenamefont {Carnimeo},
  \citenamefont {Corso}, \citenamefont {de~Gironcoli}, \citenamefont {Delugas},
  \citenamefont {DiStasio}, \citenamefont {Ferretti}, \citenamefont {Floris},
  \citenamefont {Fratesi}, \citenamefont {Fugallo}, \citenamefont {Gebauer},
  \citenamefont {Gerstmann}, \citenamefont {Giustino}, \citenamefont {Gorni},
  \citenamefont {Jia}, \citenamefont {Kawamura}, \citenamefont {Ko},
  \citenamefont {Kokalj}, \citenamefont {K\"{u}\c{c}\"{u}kbenli}, \citenamefont
  {Lazzeri}, \citenamefont {Marsili}, \citenamefont {Marzari}, \citenamefont
  {Mauri}, \citenamefont {Nguyen}, \citenamefont {Nguyen}, \citenamefont {de-la
  Roza}, \citenamefont {Paulatto}, \citenamefont {Ponc\'{e}}, \citenamefont
  {Rocca}, \citenamefont {Sabatini}, \citenamefont {Santra}, \citenamefont
  {Schlipf}, \citenamefont {Seitsonen}, \citenamefont {Smogunov}, \citenamefont
  {Timrov}, \citenamefont {Thonhauser}, \citenamefont {Umari}, \citenamefont
  {Vast}, \citenamefont {Wu},\ and\ \citenamefont {Baroni}}]{QE_2017}%
  \BibitemOpen
  \bibfield  {author} {\bibinfo {author} {\bibfnamefont {P.}~\bibnamefont
  {Giannozzi}}, \bibinfo {author} {\bibfnamefont {O.}~\bibnamefont
  {Andreussi}}, \bibinfo {author} {\bibfnamefont {T.}~\bibnamefont {Brumme}},
  \bibinfo {author} {\bibfnamefont {O.}~\bibnamefont {Bunau}}, \bibinfo
  {author} {\bibfnamefont {M.~B.}\ \bibnamefont {Nardelli}}, \bibinfo {author}
  {\bibfnamefont {M.}~\bibnamefont {Calandra}}, \bibinfo {author}
  {\bibfnamefont {R.}~\bibnamefont {Car}}, \bibinfo {author} {\bibfnamefont
  {C.}~\bibnamefont {Cavazzoni}}, \bibinfo {author} {\bibfnamefont
  {D.}~\bibnamefont {Ceresoli}}, \bibinfo {author} {\bibfnamefont
  {M.}~\bibnamefont {Cococcioni}}, \bibinfo {author} {\bibfnamefont
  {N.}~\bibnamefont {Colonna}}, \bibinfo {author} {\bibfnamefont
  {I.}~\bibnamefont {Carnimeo}}, \bibinfo {author} {\bibfnamefont {A.~D.}\
  \bibnamefont {Corso}}, \bibinfo {author} {\bibfnamefont {S.}~\bibnamefont
  {de~Gironcoli}}, \bibinfo {author} {\bibfnamefont {P.}~\bibnamefont
  {Delugas}}, \bibinfo {author} {\bibfnamefont {R.~A.}\ \bibnamefont
  {DiStasio}}, \bibinfo {author} {\bibfnamefont {A.}~\bibnamefont {Ferretti}},
  \bibinfo {author} {\bibfnamefont {A.}~\bibnamefont {Floris}}, \bibinfo
  {author} {\bibfnamefont {G.}~\bibnamefont {Fratesi}}, \bibinfo {author}
  {\bibfnamefont {G.}~\bibnamefont {Fugallo}}, \bibinfo {author} {\bibfnamefont
  {R.}~\bibnamefont {Gebauer}}, \bibinfo {author} {\bibfnamefont
  {U.}~\bibnamefont {Gerstmann}}, \bibinfo {author} {\bibfnamefont
  {F.}~\bibnamefont {Giustino}}, \bibinfo {author} {\bibfnamefont
  {T.}~\bibnamefont {Gorni}}, \bibinfo {author} {\bibfnamefont
  {J.}~\bibnamefont {Jia}}, \bibinfo {author} {\bibfnamefont {M.}~\bibnamefont
  {Kawamura}}, \bibinfo {author} {\bibfnamefont {H.-Y.}\ \bibnamefont {Ko}},
  \bibinfo {author} {\bibfnamefont {A.}~\bibnamefont {Kokalj}}, \bibinfo
  {author} {\bibfnamefont {E.}~\bibnamefont {K\"{u}\c{c}\"{u}kbenli}}, \bibinfo
  {author} {\bibfnamefont {M.}~\bibnamefont {Lazzeri}}, \bibinfo {author}
  {\bibfnamefont {M.}~\bibnamefont {Marsili}}, \bibinfo {author} {\bibfnamefont
  {N.}~\bibnamefont {Marzari}}, \bibinfo {author} {\bibfnamefont
  {F.}~\bibnamefont {Mauri}}, \bibinfo {author} {\bibfnamefont {N.~L.}\
  \bibnamefont {Nguyen}}, \bibinfo {author} {\bibfnamefont {H.-V.}\
  \bibnamefont {Nguyen}}, \bibinfo {author} {\bibfnamefont {A.~O.}\
  \bibnamefont {de-la Roza}}, \bibinfo {author} {\bibfnamefont
  {L.}~\bibnamefont {Paulatto}}, \bibinfo {author} {\bibfnamefont
  {S.}~\bibnamefont {Ponc\'{e}}}, \bibinfo {author} {\bibfnamefont
  {D.}~\bibnamefont {Rocca}}, \bibinfo {author} {\bibfnamefont
  {R.}~\bibnamefont {Sabatini}}, \bibinfo {author} {\bibfnamefont
  {B.}~\bibnamefont {Santra}}, \bibinfo {author} {\bibfnamefont
  {M.}~\bibnamefont {Schlipf}}, \bibinfo {author} {\bibfnamefont {A.~P.}\
  \bibnamefont {Seitsonen}}, \bibinfo {author} {\bibfnamefont {A.}~\bibnamefont
  {Smogunov}}, \bibinfo {author} {\bibfnamefont {I.}~\bibnamefont {Timrov}},
  \bibinfo {author} {\bibfnamefont {T.}~\bibnamefont {Thonhauser}}, \bibinfo
  {author} {\bibfnamefont {P.}~\bibnamefont {Umari}}, \bibinfo {author}
  {\bibfnamefont {N.}~\bibnamefont {Vast}}, \bibinfo {author} {\bibfnamefont
  {X.}~\bibnamefont {Wu}},\ and\ \bibinfo {author} {\bibfnamefont
  {S.}~\bibnamefont {Baroni}},\ }\href
  {https://doi.org/10.1088/1361-648X/aa8f79} {\bibfield  {journal} {\bibinfo
  {journal} {J. Phys.: Condens. Matter}\ }\textbf {\bibinfo {volume} {29}},\
  \bibinfo {pages} {465901} (\bibinfo {year} {2017})}\BibitemShut {NoStop}%
\bibitem [{\citenamefont {Giannozzi}\ \emph {et~al.}(2020)\citenamefont
  {Giannozzi}, \citenamefont {Baseggio}, \citenamefont {Bonf$\mathrm{\grave
  a}$}, \citenamefont {Brunato}, \citenamefont {Car}, \citenamefont {Carnimeo},
  \citenamefont {Cavazzoni}, \citenamefont {de~Gironcoli}, \citenamefont
  {Delugas}, \citenamefont {Ferrari~Ruffino}, \citenamefont {Ferretti},
  \citenamefont {Marzari}, \citenamefont {Timrov}, \citenamefont {Urru},\ and\
  \citenamefont {Baroni}}]{QE_2020_GPU}%
  \BibitemOpen
  \bibfield  {author} {\bibinfo {author} {\bibfnamefont {P.}~\bibnamefont
  {Giannozzi}}, \bibinfo {author} {\bibfnamefont {O.}~\bibnamefont {Baseggio}},
  \bibinfo {author} {\bibfnamefont {P.}~\bibnamefont {Bonf$\mathrm{\grave
  a}$}}, \bibinfo {author} {\bibfnamefont {D.}~\bibnamefont {Brunato}},
  \bibinfo {author} {\bibfnamefont {R.}~\bibnamefont {Car}}, \bibinfo {author}
  {\bibfnamefont {I.}~\bibnamefont {Carnimeo}}, \bibinfo {author}
  {\bibfnamefont {C.}~\bibnamefont {Cavazzoni}}, \bibinfo {author}
  {\bibfnamefont {S.}~\bibnamefont {de~Gironcoli}}, \bibinfo {author}
  {\bibfnamefont {P.}~\bibnamefont {Delugas}}, \bibinfo {author} {\bibfnamefont
  {F.}~\bibnamefont {Ferrari~Ruffino}}, \bibinfo {author} {\bibfnamefont
  {A.}~\bibnamefont {Ferretti}}, \bibinfo {author} {\bibfnamefont
  {N.}~\bibnamefont {Marzari}}, \bibinfo {author} {\bibfnamefont
  {I.}~\bibnamefont {Timrov}}, \bibinfo {author} {\bibfnamefont
  {A.}~\bibnamefont {Urru}},\ and\ \bibinfo {author} {\bibfnamefont
  {S.}~\bibnamefont {Baroni}},\ }\bibfield  {journal} {\bibinfo  {journal} {The
  Journal of Chemical Physics}\ }\textbf {\bibinfo {volume} {152}},\ \href
  {https://doi.org/10.1063/5.0005082} {10.1063/5.0005082} (\bibinfo {year}
  {2020}),\ \bibinfo {note} {154105}\BibitemShut {NoStop}%
\bibitem [{sm()}]{sm}%
  \BibitemOpen
  \href@noop {} {}\bibinfo {howpublished} {See Supplemental Material at [URL
  will be inserted by publisher] for sample characterization, details of DFT
  calculations, simulations of internal magnetic field at Mu sites, and $\mu$SR
  results for a powder sample.}\BibitemShut {Stop}%
\bibitem [{\citenamefont {Rodr^^c3^^adguez-Carvajal}(1993)}]{fullprof}%
  \BibitemOpen
  \bibfield  {author} {\bibinfo {author} {\bibfnamefont {J.}~\bibnamefont
  {Rodr^^c3^^adguez-Carvajal}},\ }\href
  {https://doi.org/https://doi.org/10.1016/0921-4526(93)90108-I} {\bibfield
  {journal} {\bibinfo  {journal} {Physica B: Condensed Matter}\ }\textbf
  {\bibinfo {volume} {192}},\ \bibinfo {pages} {55} (\bibinfo {year}
  {1993})}\BibitemShut {NoStop}%
\bibitem [{\citenamefont {Dolomanov}\ \emph {et~al.}(2009)\citenamefont
  {Dolomanov}, \citenamefont {Bourhis}, \citenamefont {Gildea}, \citenamefont
  {Howard},\ and\ \citenamefont {Puschmann}}]{olex2_1}%
  \BibitemOpen
  \bibfield  {author} {\bibinfo {author} {\bibfnamefont {O.~V.}\ \bibnamefont
  {Dolomanov}}, \bibinfo {author} {\bibfnamefont {L.~J.}\ \bibnamefont
  {Bourhis}}, \bibinfo {author} {\bibfnamefont {R.~J.}\ \bibnamefont {Gildea}},
  \bibinfo {author} {\bibfnamefont {J.~A.~K.}\ \bibnamefont {Howard}},\ and\
  \bibinfo {author} {\bibfnamefont {H.}~\bibnamefont {Puschmann}},\ }\href
  {https://doi.org/https://doi.org/10.1107/S0021889808042726} {\bibfield
  {journal} {\bibinfo  {journal} {J. Appl. Crystallogr.}\ }\textbf {\bibinfo
  {volume} {42}},\ \bibinfo {pages} {339} (\bibinfo {year} {2009})}\BibitemShut
  {NoStop}%
\bibitem [{\citenamefont {Bourhis}\ \emph {et~al.}(2015)\citenamefont
  {Bourhis}, \citenamefont {Dolomanov}, \citenamefont {Gildea}, \citenamefont
  {Howard},\ and\ \citenamefont {Puschmann}}]{olex2_2}%
  \BibitemOpen
  \bibfield  {author} {\bibinfo {author} {\bibfnamefont {L.~J.}\ \bibnamefont
  {Bourhis}}, \bibinfo {author} {\bibfnamefont {O.~V.}\ \bibnamefont
  {Dolomanov}}, \bibinfo {author} {\bibfnamefont {R.~J.}\ \bibnamefont
  {Gildea}}, \bibinfo {author} {\bibfnamefont {J.~A.}\ \bibnamefont {Howard}},\
  and\ \bibinfo {author} {\bibfnamefont {H.}~\bibnamefont {Puschmann}},\
  }\href@noop {} {\bibfield  {journal} {\bibinfo  {journal} {Acta Crystallogr A
  Found Adv}\ }\textbf {\bibinfo {volume} {71}},\ \bibinfo {pages} {59}
  (\bibinfo {year} {2015})}\BibitemShut {NoStop}%
\bibitem [{\citenamefont {Sheldrick}(2015)}]{Shelxl}%
  \BibitemOpen
  \bibfield  {author} {\bibinfo {author} {\bibfnamefont {G.~M.}\ \bibnamefont
  {Sheldrick}},\ }\href {https://doi.org/10.1107/S2053229614024218} {\bibfield
  {journal} {\bibinfo  {journal} {Acta Crystallographica Section C}\ }\textbf
  {\bibinfo {volume} {71}},\ \bibinfo {pages} {3} (\bibinfo {year}
  {2015})}\BibitemShut {NoStop}%
\bibitem [{PP()}]{PP}%
  \BibitemOpen
  \href {https://dalcorso.github.io/pslibrary} {}\bibinfo {howpublished}
  {https://dalcorso.github.io/pslibrary}\BibitemShut {NoStop}%
\bibitem [{\citenamefont {Hamann}(2013)}]{ONCV}%
  \BibitemOpen
  \bibfield  {author} {\bibinfo {author} {\bibfnamefont {D.~R.}\ \bibnamefont
  {Hamann}},\ }\href {https://doi.org/10.1103/PhysRevB.88.085117} {\bibfield
  {journal} {\bibinfo  {journal} {Phys. Rev. B}\ }\textbf {\bibinfo {volume}
  {88}},\ \bibinfo {pages} {085117} (\bibinfo {year} {2013})}\BibitemShut
  {NoStop}%
\bibitem [{\citenamefont {Rogers}\ \emph {et~al.}(1969)\citenamefont {Rogers},
  \citenamefont {Shannon}, \citenamefont {Sleight},\ and\ \citenamefont
  {Gillson}}]{Rogers:69}%
  \BibitemOpen
  \bibfield  {author} {\bibinfo {author} {\bibfnamefont {D.~B.}\ \bibnamefont
  {Rogers}}, \bibinfo {author} {\bibfnamefont {R.~D.}\ \bibnamefont {Shannon}},
  \bibinfo {author} {\bibfnamefont {A.~W.}\ \bibnamefont {Sleight}},\ and\
  \bibinfo {author} {\bibfnamefont {J.~L.}\ \bibnamefont {Gillson}},\ }\href
  {https://doi.org/10.1021/ic50074a029} {\bibfield  {journal} {\bibinfo
  {journal} {Inorg. Chem.}\ }\textbf {\bibinfo {volume} {8}},\ \bibinfo {pages}
  {841} (\bibinfo {year} {1969})}\BibitemShut {NoStop}%
\bibitem [{\citenamefont {Kojima}\ \emph {et~al.}(2014)\citenamefont {Kojima},
  \citenamefont {Murakami}, \citenamefont {Takahashi}, \citenamefont {Lee},
  \citenamefont {Suzuki}, \citenamefont {Koda}, \citenamefont {Yamauchi},
  \citenamefont {Miyazaki}, \citenamefont {Hiraishi}, \citenamefont {Okabe},
  \citenamefont {Takeshita}, \citenamefont {Kadono}, \citenamefont {Ito},
  \citenamefont {Higemoto}, \citenamefont {Kanda}, \citenamefont {Fukao},
  \citenamefont {Saito}, \citenamefont {Saito}, \citenamefont {Ikeno},
  \citenamefont {Uchida},\ and\ \citenamefont {Tanaka}}]{ARTEMIS}%
  \BibitemOpen
  \bibfield  {author} {\bibinfo {author} {\bibfnamefont {K.~M.}\ \bibnamefont
  {Kojima}}, \bibinfo {author} {\bibfnamefont {T.}~\bibnamefont {Murakami}},
  \bibinfo {author} {\bibfnamefont {Y.}~\bibnamefont {Takahashi}}, \bibinfo
  {author} {\bibfnamefont {H.}~\bibnamefont {Lee}}, \bibinfo {author}
  {\bibfnamefont {S.~Y.}\ \bibnamefont {Suzuki}}, \bibinfo {author}
  {\bibfnamefont {A.}~\bibnamefont {Koda}}, \bibinfo {author} {\bibfnamefont
  {I.}~\bibnamefont {Yamauchi}}, \bibinfo {author} {\bibfnamefont
  {M.}~\bibnamefont {Miyazaki}}, \bibinfo {author} {\bibfnamefont
  {M.}~\bibnamefont {Hiraishi}}, \bibinfo {author} {\bibfnamefont
  {H.}~\bibnamefont {Okabe}}, \bibinfo {author} {\bibfnamefont
  {S.}~\bibnamefont {Takeshita}}, \bibinfo {author} {\bibfnamefont
  {R.}~\bibnamefont {Kadono}}, \bibinfo {author} {\bibfnamefont
  {T.}~\bibnamefont {Ito}}, \bibinfo {author} {\bibfnamefont {W.}~\bibnamefont
  {Higemoto}}, \bibinfo {author} {\bibfnamefont {S.}~\bibnamefont {Kanda}},
  \bibinfo {author} {\bibfnamefont {Y.}~\bibnamefont {Fukao}}, \bibinfo
  {author} {\bibfnamefont {N.}~\bibnamefont {Saito}}, \bibinfo {author}
  {\bibfnamefont {M.}~\bibnamefont {Saito}}, \bibinfo {author} {\bibfnamefont
  {M.}~\bibnamefont {Ikeno}}, \bibinfo {author} {\bibfnamefont
  {T.}~\bibnamefont {Uchida}},\ and\ \bibinfo {author} {\bibfnamefont {M.~M.}\
  \bibnamefont {Tanaka}},\ }\href
  {https://doi.org/10.1088/1742-6596/551/1/012063} {\bibfield  {journal}
  {\bibinfo  {journal} {J. Phys.: Conf. Ser.}\ }\textbf {\bibinfo {volume}
  {551}},\ \bibinfo {pages} {012063} (\bibinfo {year} {2014})}\BibitemShut
  {NoStop}%
\bibitem [{\citenamefont {Suter}\ and\ \citenamefont {Wojek}(2012)}]{musrfit}%
  \BibitemOpen
  \bibfield  {author} {\bibinfo {author} {\bibfnamefont {A.}~\bibnamefont
  {Suter}}\ and\ \bibinfo {author} {\bibfnamefont {B.}~\bibnamefont {Wojek}},\
  }\href {https://doi.org/https://doi.org/10.1016/j.phpro.2012.04.042}
  {\bibfield  {journal} {\bibinfo  {journal} {Phys. Proc.}\ }\textbf {\bibinfo
  {volume} {30}},\ \bibinfo {pages} {69} (\bibinfo {year} {2012})}\BibitemShut
  {NoStop}%
\bibitem [{\citenamefont {Kojima}()}]{dipelec}%
  \BibitemOpen
  \bibfield  {author} {\bibinfo {author} {\bibfnamefont {K.~M.}\ \bibnamefont
  {Kojima}},\ }\href@noop {} {}\bibinfo {howpublished} {private
  communication}\BibitemShut {NoStop}%
\bibitem [{\citenamefont {Uemura}\ \emph {et~al.}(1985)\citenamefont {Uemura},
  \citenamefont {Yamazaki}, \citenamefont {Harshman}, \citenamefont {Senba},\
  and\ \citenamefont {Ansaldo}}]{Uemura_1985}%
  \BibitemOpen
  \bibfield  {author} {\bibinfo {author} {\bibfnamefont {Y.~J.}\ \bibnamefont
  {Uemura}}, \bibinfo {author} {\bibfnamefont {T.}~\bibnamefont {Yamazaki}},
  \bibinfo {author} {\bibfnamefont {D.~R.}\ \bibnamefont {Harshman}}, \bibinfo
  {author} {\bibfnamefont {M.}~\bibnamefont {Senba}},\ and\ \bibinfo {author}
  {\bibfnamefont {E.~J.}\ \bibnamefont {Ansaldo}},\ }\href
  {https://doi.org/10.1103/PhysRevB.31.546} {\bibfield  {journal} {\bibinfo
  {journal} {Phys. Rev. B}\ }\textbf {\bibinfo {volume} {31}},\ \bibinfo
  {pages} {546} (\bibinfo {year} {1985})}\BibitemShut {NoStop}%
\bibitem [{\citenamefont {Hayano}\ \emph {et~al.}(1979)\citenamefont {Hayano},
  \citenamefont {Uemura}, \citenamefont {Imazato}, \citenamefont {Nishida},
  \citenamefont {Yamazaki},\ and\ \citenamefont {Kubo}}]{KT}%
  \BibitemOpen
  \bibfield  {author} {\bibinfo {author} {\bibfnamefont {R.~S.}\ \bibnamefont
  {Hayano}}, \bibinfo {author} {\bibfnamefont {Y.~J.}\ \bibnamefont {Uemura}},
  \bibinfo {author} {\bibfnamefont {J.}~\bibnamefont {Imazato}}, \bibinfo
  {author} {\bibfnamefont {N.}~\bibnamefont {Nishida}}, \bibinfo {author}
  {\bibfnamefont {T.}~\bibnamefont {Yamazaki}},\ and\ \bibinfo {author}
  {\bibfnamefont {R.}~\bibnamefont {Kubo}},\ }\href
  {https://doi.org/10.1103/PhysRevB.20.850} {\bibfield  {journal} {\bibinfo
  {journal} {Phys. Rev. B}\ }\textbf {\bibinfo {volume} {20}},\ \bibinfo
  {pages} {850} (\bibinfo {year} {1979})}\BibitemShut {NoStop}%
\bibitem [{\citenamefont {Cox}\ \emph {et~al.}(2006)\citenamefont {Cox},
  \citenamefont {Gavartin}, \citenamefont {Lord}, \citenamefont {Cottrell},
  \citenamefont {Gil}, \citenamefont {Alberto}, \citenamefont {Duarte},
  \citenamefont {Vil$\mathrm{\tilde a}$o}, \citenamefont {de~Campos},
  \citenamefont {Keeble}, \citenamefont {Davis}, \citenamefont {Charlton},\
  and\ \citenamefont {van~der Werf}}]{Cox_06}%
  \BibitemOpen
  \bibfield  {author} {\bibinfo {author} {\bibfnamefont {S.~F.~J.}\
  \bibnamefont {Cox}}, \bibinfo {author} {\bibfnamefont {J.~L.}\ \bibnamefont
  {Gavartin}}, \bibinfo {author} {\bibfnamefont {J.~S.}\ \bibnamefont {Lord}},
  \bibinfo {author} {\bibfnamefont {S.~P.}\ \bibnamefont {Cottrell}}, \bibinfo
  {author} {\bibfnamefont {J.~M.}\ \bibnamefont {Gil}}, \bibinfo {author}
  {\bibfnamefont {H.~V.}\ \bibnamefont {Alberto}}, \bibinfo {author}
  {\bibfnamefont {J.~P.}\ \bibnamefont {Duarte}}, \bibinfo {author}
  {\bibfnamefont {R.~C.}\ \bibnamefont {Vil$\mathrm{\tilde a}$o}}, \bibinfo
  {author} {\bibfnamefont {N.~A.}\ \bibnamefont {de~Campos}}, \bibinfo {author}
  {\bibfnamefont {D.~J.}\ \bibnamefont {Keeble}}, \bibinfo {author}
  {\bibfnamefont {E.~A.}\ \bibnamefont {Davis}}, \bibinfo {author}
  {\bibfnamefont {M.}~\bibnamefont {Charlton}},\ and\ \bibinfo {author}
  {\bibfnamefont {D.~P.}\ \bibnamefont {van~der Werf}},\ }\href
  {https://doi.org/10.1088/0953-8984/18/3/022} {\bibfield  {journal} {\bibinfo
  {journal} {J. Phys.: Condens. Matter}\ }\textbf {\bibinfo {volume} {18}},\
  \bibinfo {pages} {1079} (\bibinfo {year} {2006})}\BibitemShut {NoStop}%
\bibitem [{\citenamefont {Ito}\ \emph {et~al.}(2023)\citenamefont {Ito},
  \citenamefont {Higemoto},\ and\ \citenamefont {Shimomura}}]{TUIto_23}%
  \BibitemOpen
  \bibfield  {author} {\bibinfo {author} {\bibfnamefont {T.~U.}\ \bibnamefont
  {Ito}}, \bibinfo {author} {\bibfnamefont {W.}~\bibnamefont {Higemoto}},\ and\
  \bibinfo {author} {\bibfnamefont {K.}~\bibnamefont {Shimomura}},\ }\href
  {https://doi.org/10.1103/PhysRevB.108.224301} {\bibfield  {journal} {\bibinfo
   {journal} {Phys. Rev. B}\ }\textbf {\bibinfo {volume} {108}},\ \bibinfo
  {pages} {224301} (\bibinfo {year} {2023})}\BibitemShut {NoStop}%
\bibitem [{\citenamefont {Jadon}\ \emph {et~al.}(2021)\citenamefont {Jadon},
  \citenamefont {Rossi}, \citenamefont {Djafari-Rouhani}, \citenamefont
  {Est$\mathrm{\grave e}$ve},\ and\ \citenamefont {Pech}}]{RuO2_NEB_2021}%
  \BibitemOpen
  \bibfield  {author} {\bibinfo {author} {\bibfnamefont {A.}~\bibnamefont
  {Jadon}}, \bibinfo {author} {\bibfnamefont {C.}~\bibnamefont {Rossi}},
  \bibinfo {author} {\bibfnamefont {M.}~\bibnamefont {Djafari-Rouhani}},
  \bibinfo {author} {\bibfnamefont {A.}~\bibnamefont {Est$\mathrm{\grave
  e}$ve}},\ and\ \bibinfo {author} {\bibfnamefont {D.}~\bibnamefont {Pech}},\
  }\href {https://doi.org/https://doi.org/10.1016/j.physo.2021.100059}
  {\bibfield  {journal} {\bibinfo  {journal} {Physics Open}\ }\textbf {\bibinfo
  {volume} {7}},\ \bibinfo {pages} {100059} (\bibinfo {year}
  {2021})}\BibitemShut {NoStop}%
\bibitem [{\citenamefont {Marinopoulos}\ \emph {et~al.}(2018)\citenamefont
  {Marinopoulos}, \citenamefont {Vil$\mathrm{\tilde a}$o}, \citenamefont
  {Alberto},\ and\ \citenamefont {Gil}}]{TiO2_NEB_2018}%
  \BibitemOpen
  \bibfield  {author} {\bibinfo {author} {\bibfnamefont {A.~G.}\ \bibnamefont
  {Marinopoulos}}, \bibinfo {author} {\bibfnamefont {R.~C.}\ \bibnamefont
  {Vil$\mathrm{\tilde a}$o}}, \bibinfo {author} {\bibfnamefont {H.~V.}\
  \bibnamefont {Alberto}},\ and\ \bibinfo {author} {\bibfnamefont {J.~M.}\
  \bibnamefont {Gil}},\ }\href {https://doi.org/10.1088/1361-648X/aae0a2}
  {\bibfield  {journal} {\bibinfo  {journal} {J. Phys.: Condens. Matter}\
  }\textbf {\bibinfo {volume} {30}},\ \bibinfo {pages} {425503} (\bibinfo
  {year} {2018})}\BibitemShut {NoStop}%
\bibitem [{\citenamefont {Hiraishi}\ \emph {et~al.}(2020)\citenamefont
  {Hiraishi}, \citenamefont {Kojima}, \citenamefont {Okabe}, \citenamefont
  {Takeshita}, \citenamefont {Koda}, \citenamefont {Kadono}, \citenamefont
  {Khasanov}, \citenamefont {Iimura}, \citenamefont {Matsuishi},\ and\
  \citenamefont {Hosono}}]{MH_PRB20}%
  \BibitemOpen
  \bibfield  {author} {\bibinfo {author} {\bibfnamefont {M.}~\bibnamefont
  {Hiraishi}}, \bibinfo {author} {\bibfnamefont {K.~M.}\ \bibnamefont
  {Kojima}}, \bibinfo {author} {\bibfnamefont {H.}~\bibnamefont {Okabe}},
  \bibinfo {author} {\bibfnamefont {S.}~\bibnamefont {Takeshita}}, \bibinfo
  {author} {\bibfnamefont {A.}~\bibnamefont {Koda}}, \bibinfo {author}
  {\bibfnamefont {R.}~\bibnamefont {Kadono}}, \bibinfo {author} {\bibfnamefont
  {R.}~\bibnamefont {Khasanov}}, \bibinfo {author} {\bibfnamefont
  {S.}~\bibnamefont {Iimura}}, \bibinfo {author} {\bibfnamefont
  {S.}~\bibnamefont {Matsuishi}},\ and\ \bibinfo {author} {\bibfnamefont
  {H.}~\bibnamefont {Hosono}},\ }\href
  {https://doi.org/10.1103/PhysRevB.101.174414} {\bibfield  {journal} {\bibinfo
   {journal} {Phys. Rev. B}\ }\textbf {\bibinfo {volume} {101}},\ \bibinfo
  {pages} {174414} (\bibinfo {year} {2020})}\BibitemShut {NoStop}%
\bibitem [{\citenamefont {Kim}\ and\ \citenamefont {Lai}(2015)}]{NEB_2015}%
  \BibitemOpen
  \bibfield  {author} {\bibinfo {author} {\bibfnamefont {S.}~\bibnamefont
  {Kim}}\ and\ \bibinfo {author} {\bibfnamefont {W.}~\bibnamefont {Lai}},\
  }\href {https://doi.org/https://doi.org/10.1016/j.commatsci.2015.01.016}
  {\bibfield  {journal} {\bibinfo  {journal} {Comput. Mater. Sci.}\ }\textbf
  {\bibinfo {volume} {102}},\ \bibinfo {pages} {14} (\bibinfo {year}
  {2015})}\BibitemShut {NoStop}%
\bibitem [{\citenamefont {Momma}\ and\ \citenamefont {Izumi}(2011)}]{Vesta}%
  \BibitemOpen
  \bibfield  {author} {\bibinfo {author} {\bibfnamefont {K.}~\bibnamefont
  {Momma}}\ and\ \bibinfo {author} {\bibfnamefont {F.}~\bibnamefont {Izumi}},\
  }\href {https://doi.org/10.1107/S0021889811038970} {\bibfield  {journal}
  {\bibinfo  {journal} {J. Appl. Crystallogr.}\ }\textbf {\bibinfo {volume}
  {44}},\ \bibinfo {pages} {1272} (\bibinfo {year} {2011})}\BibitemShut
  {NoStop}%
\bibitem [{\citenamefont {Yamaura}\ \emph {et~al.}(2019)\citenamefont
  {Yamaura}, \citenamefont {Hiraka}, \citenamefont {Iimura}, \citenamefont
  {Muraba}, \citenamefont {Bang}, \citenamefont {Ikeuchi}, \citenamefont
  {Nakamura}, \citenamefont {Inamura}, \citenamefont {Honda}, \citenamefont
  {Hiraishi}, \citenamefont {Kojima}, \citenamefont {Kadono}, \citenamefont
  {Kuramoto}, \citenamefont {Murakami}, \citenamefont {Matsuishi},\ and\
  \citenamefont {Hosono}}]{Yamaura_2019}%
  \BibitemOpen
  \bibfield  {author} {\bibinfo {author} {\bibfnamefont {J.-i.}\ \bibnamefont
  {Yamaura}}, \bibinfo {author} {\bibfnamefont {H.}~\bibnamefont {Hiraka}},
  \bibinfo {author} {\bibfnamefont {S.}~\bibnamefont {Iimura}}, \bibinfo
  {author} {\bibfnamefont {Y.}~\bibnamefont {Muraba}}, \bibinfo {author}
  {\bibfnamefont {J.}~\bibnamefont {Bang}}, \bibinfo {author} {\bibfnamefont
  {K.}~\bibnamefont {Ikeuchi}}, \bibinfo {author} {\bibfnamefont
  {M.}~\bibnamefont {Nakamura}}, \bibinfo {author} {\bibfnamefont
  {Y.}~\bibnamefont {Inamura}}, \bibinfo {author} {\bibfnamefont
  {T.}~\bibnamefont {Honda}}, \bibinfo {author} {\bibfnamefont
  {M.}~\bibnamefont {Hiraishi}}, \bibinfo {author} {\bibfnamefont {K.~M.}\
  \bibnamefont {Kojima}}, \bibinfo {author} {\bibfnamefont {R.}~\bibnamefont
  {Kadono}}, \bibinfo {author} {\bibfnamefont {Y.}~\bibnamefont {Kuramoto}},
  \bibinfo {author} {\bibfnamefont {Y.}~\bibnamefont {Murakami}}, \bibinfo
  {author} {\bibfnamefont {S.}~\bibnamefont {Matsuishi}},\ and\ \bibinfo
  {author} {\bibfnamefont {H.}~\bibnamefont {Hosono}},\ }\href
  {https://doi.org/10.1103/PhysRevB.99.220505} {\bibfield  {journal} {\bibinfo
  {journal} {Phys. Rev. B}\ }\textbf {\bibinfo {volume} {99}},\ \bibinfo
  {pages} {220505} (\bibinfo {year} {2019})}\BibitemShut {NoStop}%
\bibitem [{\citenamefont {Bloembergen}\ \emph {et~al.}(1948)\citenamefont
  {Bloembergen}, \citenamefont {Purcell},\ and\ \citenamefont {Pound}}]{BPP}%
  \BibitemOpen
  \bibfield  {author} {\bibinfo {author} {\bibfnamefont {N.}~\bibnamefont
  {Bloembergen}}, \bibinfo {author} {\bibfnamefont {E.~M.}\ \bibnamefont
  {Purcell}},\ and\ \bibinfo {author} {\bibfnamefont {R.~V.}\ \bibnamefont
  {Pound}},\ }\href {https://doi.org/10.1103/PhysRev.73.679} {\bibfield
  {journal} {\bibinfo  {journal} {Phys. Rev.}\ }\textbf {\bibinfo {volume}
  {73}},\ \bibinfo {pages} {679} (\bibinfo {year} {1948})}\BibitemShut
  {NoStop}%
\end{thebibliography}%
\end{document}